\documentclass{aastex631}

\shortauthors{Humphries}
\usepackage{amstext}
\usepackage{tikz}
\usepackage{booktabs}
\usepackage{graphicx}
\usepackage{epstopdf}
\usepackage{rotating}
\usepackage{amssymb}

\newcommand{\nfrag}{$N_{frag}$}

\newcommand{\paperi}{Paper I}

\newcommand{\alf}{Alfv\'en}

\begin{document}

\title{The evolution of chromospheric quiet-sun bright points during their lifetime}

\correspondingauthor{Humphries, Morgan, Kuridze}
\email{ldh6@aber.ac.uk, hum2@aber.ac.uk, dak21@aber.ac.uk}

\author[0000-0002-0786-7307]{Ll\^yr Dafydd Humphries}
\affiliation{Aberystwyth University \\
Faculty of Business and Physical Sciences\\
Aberystwyth, Ceredigion, SY23 3FL, Wales, UK}

\author[0000-0002-0786-7307]{Huw Morgan}
\affiliation{Aberystwyth University \\
Faculty of Business and Physical Sciences\\
Aberystwyth, Ceredigion, SY23 3FL, Wales, UK}

\author[0000-0003-2760-2311]{David Kuridze}
\affiliation{National Solar Observatory, 3665 Discovery Drive, Boulder, CO 80303, USA}
\affiliation{Aberystwyth University \\
Faculty of Business and Physical Sciences\\
Aberystwyth, Ceredigion, SY23 3FL, Wales, UK}
\affiliation{Georgian National Astrophysical Observatory, Abastumani, 0301, Georgia}

\begin{abstract}

Bright points (BPs) are ubiquitous, small-scale energetic events with a multithermal nature, typically observed in the chromosphere and closely linked to both the photospheric structure and the composition of the corona. Their evolution throughout their lifetimes is influenced by various physical processes, including plasma dynamics and magnetic interactions. This paper aims to gain information on the evolution of BPs through a large statistical sample, with a focus on identifying when they reach the maximum values of various attributes, and investigates whether there are statistical differences between BPs in the ''Active Quiet Sun" (AQS, above the network) and ''True Quiet Sun" (TQS, above the internetwork). The observed attributes are maximum brightness (both total and intrinsic), plane-of-sky (POS) speed, POS travel distance, POS acceleration, and apparent size (POS area). BPs can reach their maximum brightness and size at nearly any point during their lifetime, and this is likely the result of complex interactions between the BP and the surrounding chromosphere. AQS and TQS BPs are similar in most attributes and are most likely to reach their maximum POS speed halfway through their lifetimes. Positive acceleration is most likely to occur near the beginning of BP lifetimes, while negative acceleration is found to occur most commonly toward the end of BP lifetimes. Preliminary results suggest the existence of two distinct BP regimes with differential relationships between intrinsic brightness and area. We interpret these patterns as evidence that BPs, whether representing plasma motion or a propagating heating event, follow arched trajectories most likely along small magnetic loops. In this scenario, the halfway point of a BP’s lifetime corresponds to the crest of the arch, facilitating the greatest POS travel speeds. Furthermore, magnetic reconnection may occur at a photospheric footpoint, after which the BP moves along a loop before eventually returning to the photosphere at another footpoint. Deviations from expected loop-related BP evolution are likely due to complex interactions between the BP and the surrounding chromosphere and/or related to those BPs whose POS motions deviate significantly from straight lines. 
    
\end{abstract}

\section{Introduction}

The solar chromosphere and photosphere are host to a wide array of small-scale transient phenomena. These phenomena, which include spicules \citep{Beckers_1968, de_Pontieu_2007}, fibrils \citep{Rouppe_van_der_Voort_2007}, bright points \cite[BPs,][]{Berger_1996, Chitta_2012}, Ellerman bombs \citep{Ellerman_1917, Rutten_2013}, campfires \cite{Zhukov_2021}, and UV bursts \citep{Peter_2014}, exhibit complex evolution and interactions with the surrounding plasma. They are believed to play a crucial role in transporting energy through the solar atmosphere, potentially contributing to chromospheric and coronal heating \citep{Carlsson_2019, De_Pontieu_2022}.

Advances in high-resolution observations and numerical simulations have revealed that many of these events are driven by small-scale magnetic reconnection \citep{Hansteen_2019, Litvinenko_2009}, \alf\ waves \citep{Martinez-Sykora_2017}, and plasma instabilities \citep{Takasao_2013, Rouppe_van_der_Voort_2017}. The Interface Region Imaging Spectrograph \citep[IRIS,][]{de_pontieu_2014} has been instrumental in uncovering the fine-scale structure and dynamics of these events, providing unprecedented spatial resolution. IRIS has revealed details of small-scale heating events such as UV bursts \citep{Tian_2016, Young_2018} and the complex evolution of spicules \citep{Skogsrud_2015, Samanta_2019} and bright points \citep{Humphries_2021_a}, offering new insights into the mechanisms of energy transfer between the lower and upper solar atmosphere.

Despite these advances, significant uncertainties remain regarding their formation mechanisms, lifetimes, evolution, and interactions with surrounding plasma and magnetic fields \citep{Nelson_2023}. This study builds on the investigation conducted in \cite{Humphries_2024}, indeed, we use the same dataset albeit with additional constraints. 

In this work, we analyze the statistical evolution of BP attributes throughout their lifetime, with a particular focus on the time at which they reach a maximum value in certain observables. BPs detected in both ''Active" and ''Quiet" domains are directly compared. These BPs are detected, tracked, and recorded using \cite{Humphries_2021_a}'s detection method.
By examining more than 1300 detections, this analysis establishes constraints on their possible origins and the physical mechanisms driving them, while also providing additional context for how they may interact with both the photosphere and the chromosphere.  

Section \ref{sec:obs_and_meth} provides a brief summary of the observations and region of interest (ROI), as well as an overview of the analysis method. Results are presented and discussed in section \ref{sec:results}, and conclusions are provided in section \ref{sec:con}.

\begin{figure*}[t]
    \centering
    \graphicspath{{./Images}}
    \includegraphics[width=0.6\textwidth]{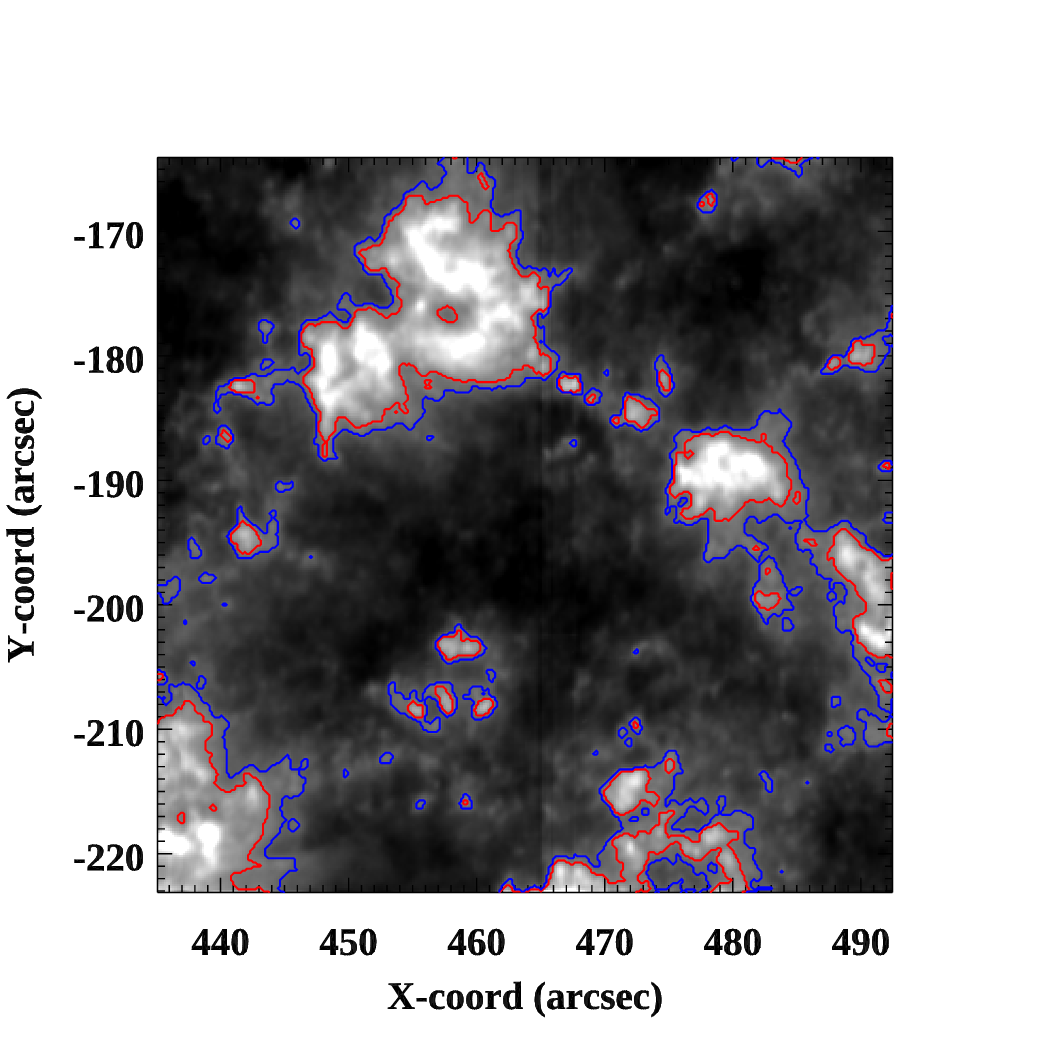}
    \caption{IRIS 1400 \AA\ image of the 2013-09-26 ROI with domain contours. Domains that lie within red contours are considered ``Active" QS. Domains outside the blue contours are considered ``True" QS. Domains that lie outside both contour colours (and any BP detections there-out) are ignored.}
    \label{fig:FOV_w_contours}
\end{figure*}

\section{Observations \& Method} \label{sec:obs_and_meth}

\begin{figure*}[t]
    \centering
    (a)\includegraphics[width=0.4\textwidth]{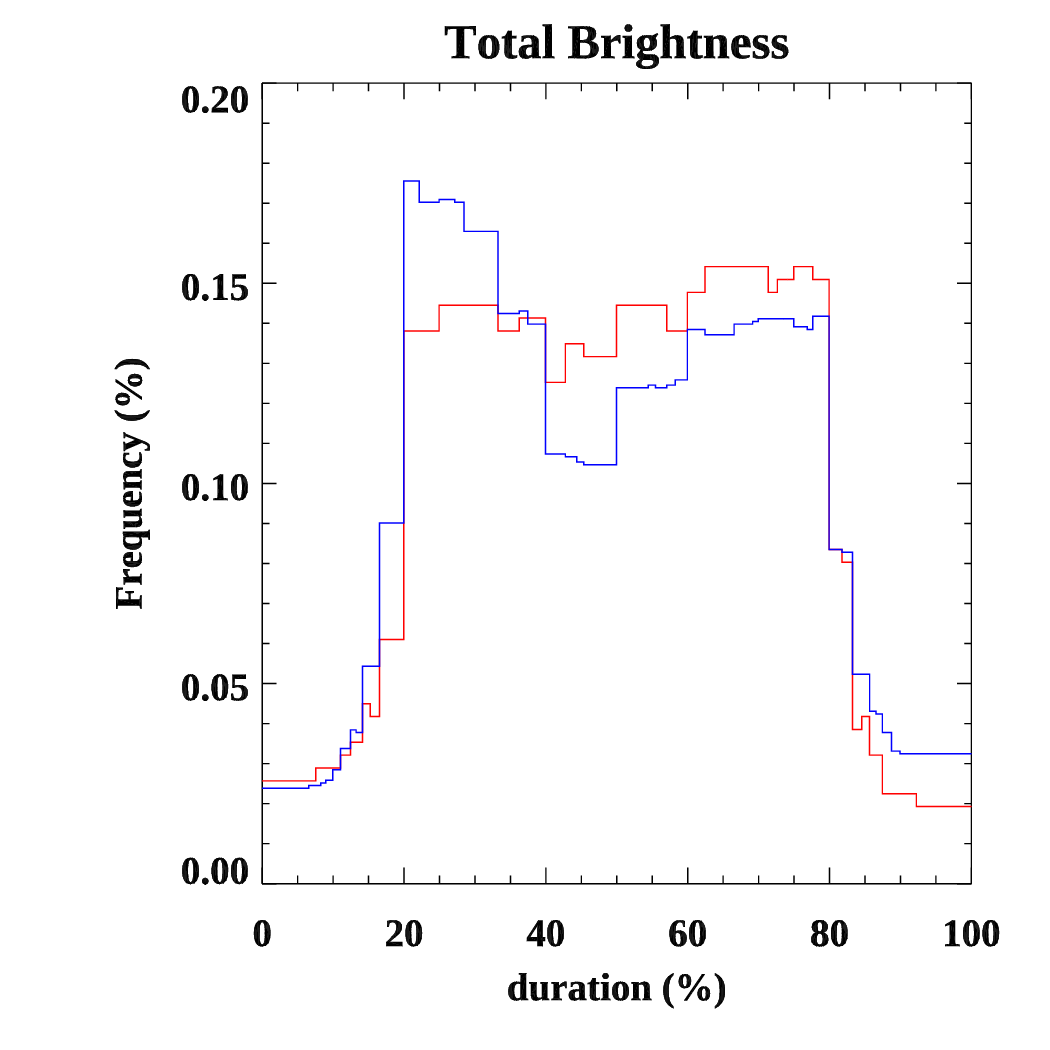}
    (b)\includegraphics[width=0.4\textwidth]{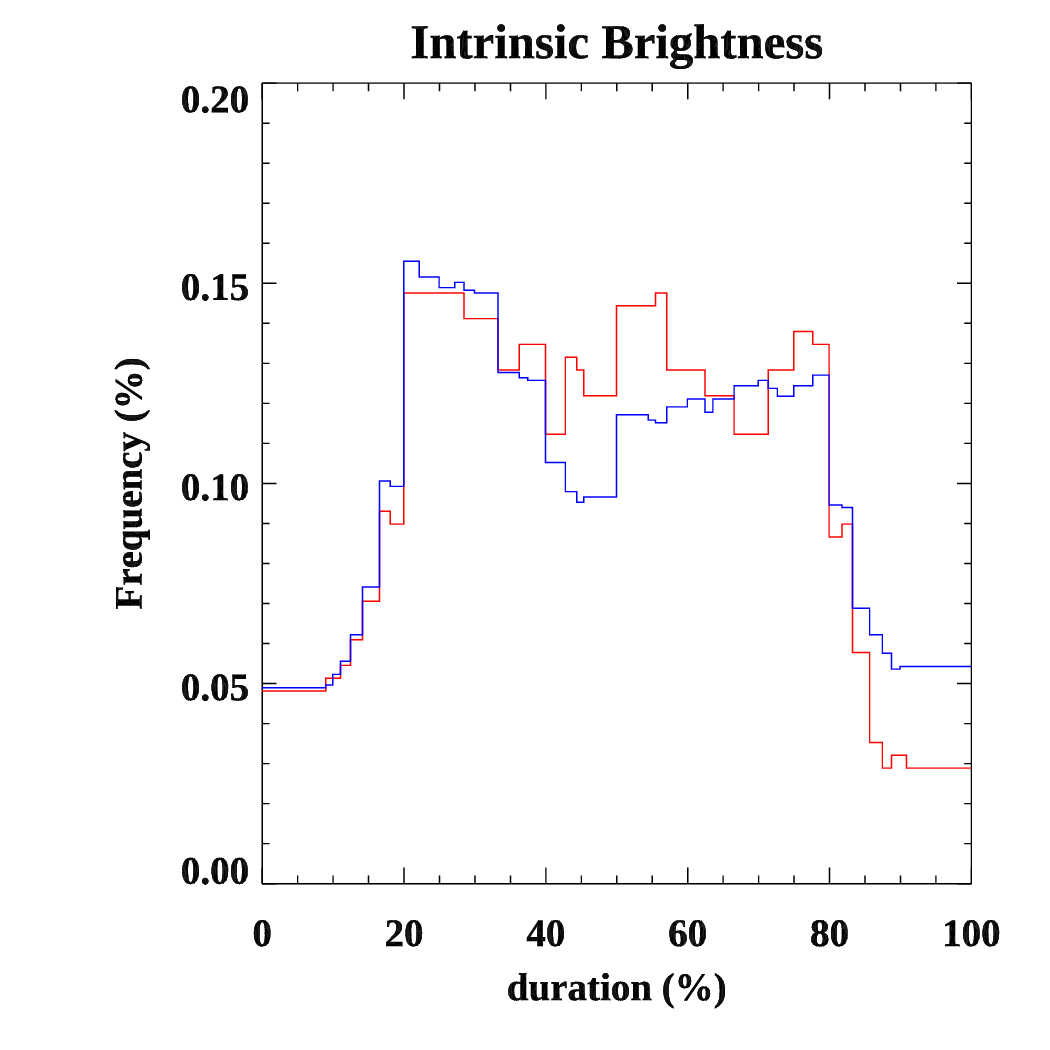}
    (c)\includegraphics[width=0.4\textwidth]{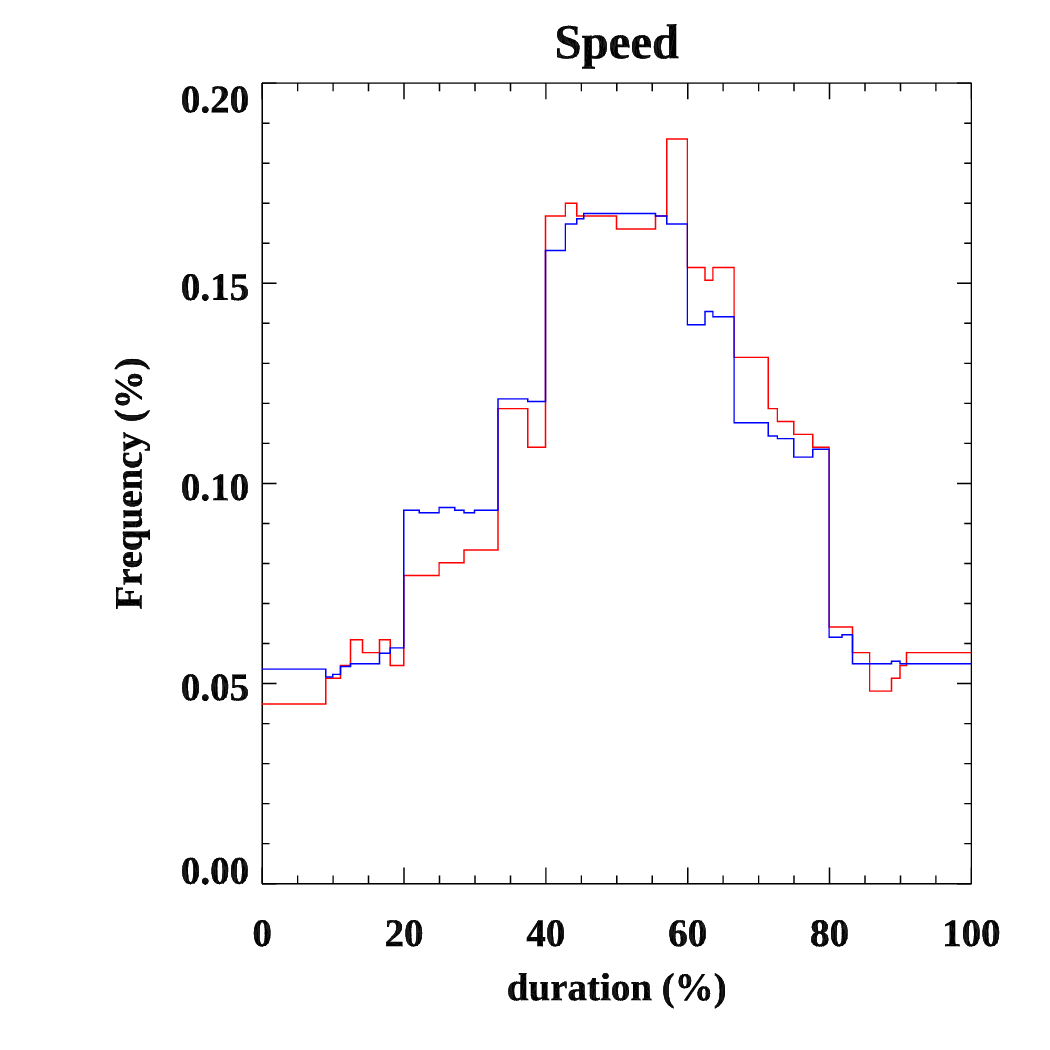}
    (d)\includegraphics[width=0.4\textwidth]{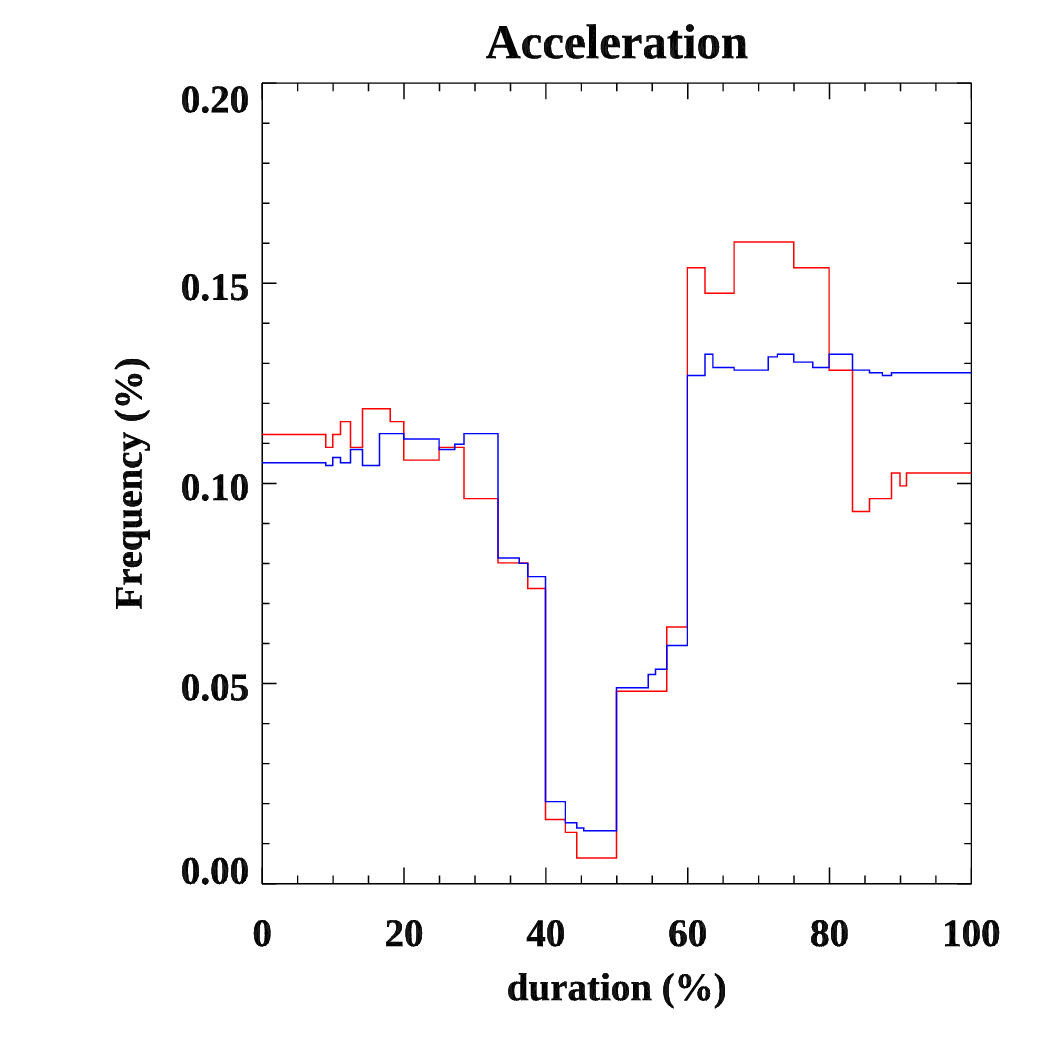}
    (e)\includegraphics[width=0.4\textwidth]{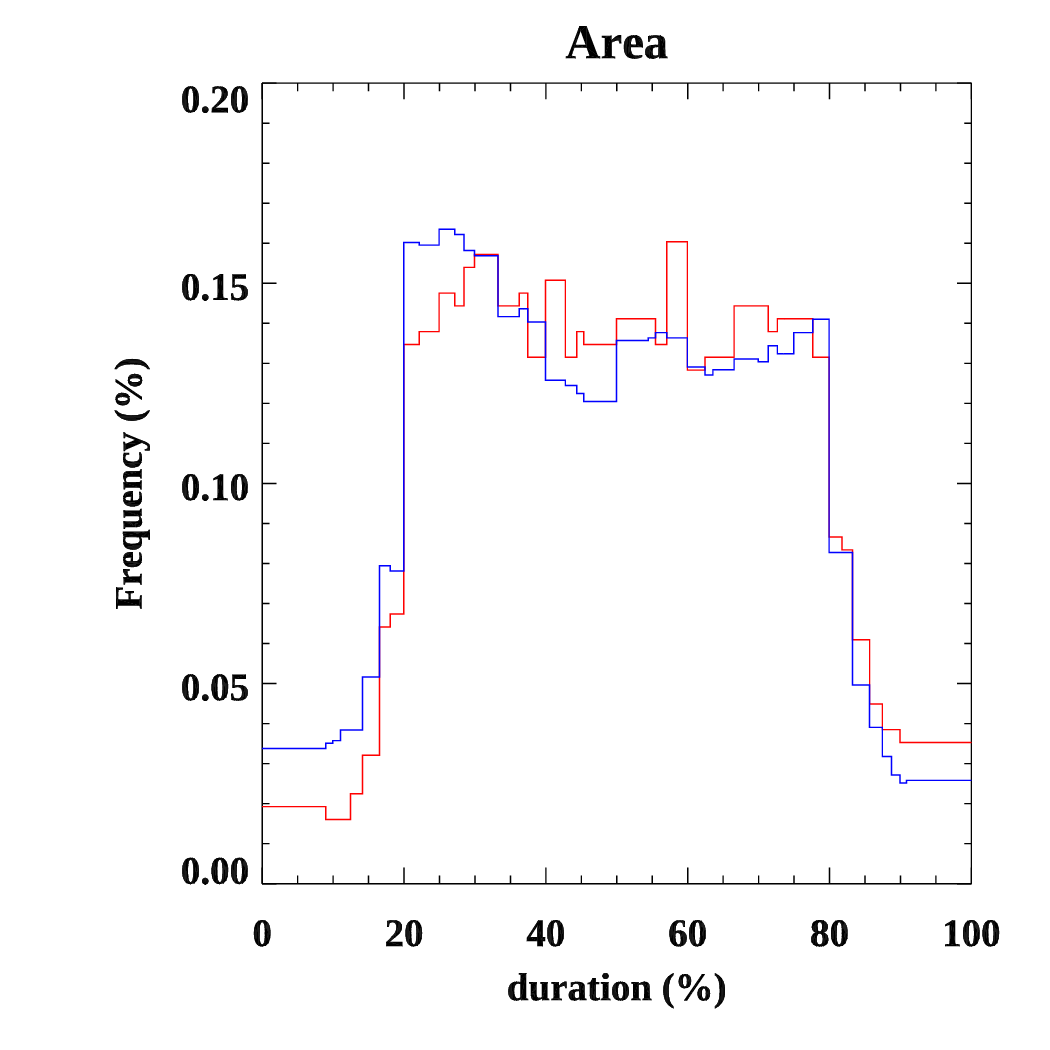}
    \label{fig:max_temp_pos}
    \caption{Histogram temporal comparison of when (a) total brightness, (b) intrinsic brigtness, (c) POS speed, (d) acceleration, and (e) area reach maximum values. Durations percentages are interpolated from differential BP durations, and only include  Nfrag=1 and duration$\le5$ results.}
\end{figure*}

\subsection{Observations}\label{sec:sub_obs}
We select an IRIS 1400 \AA\ slit-jaw image (SJI) dataset, consisting of a series of sit-and-stare images with a 0.17$\arcsec \times 0.17\arcsec$ pixel spatial scale and a nearly uniform temporal cadence of $\sim27$s, which can be seen in figure \ref{fig:FOV_w_contours}. This is the same dataset selected in \cite{Humphries_2024} and is available on the Lockheed Martin Solar and Astrophysics Laboratory (LMSAL) archive. The dataset is centred at $X = 458\arcsec$, $Y = -194\arcsec$, with a $65\arcsec\times69\arcsec$ FOV, while the cropped image of figure \ref{fig:FOV_w_contours} - is centred at $X = 464\arcsec$, $Y = -194\arcsec$, with a $57\arcsec\times59\arcsec$ FOV, having removed empty margins typical of IRIS datasets. 
The observations span from Sept. 26th 2013 at 21:26 UT to Sept. 27th 2013 at 00:37 UT, wheras the full LMSAL dataset begins earlier at 20:09 UT. However, due to contamination from high-energy particle bombardment associated with the South Atlantic Anomaly (SAA), the first $\sim170$ frames are excluded from analysis.

Similarly to \cite{Humphries_2024}, this study specifically focuses on the Si {\sc{iv}} 1400 \AA\ channel for several reasons: it is the most widely available sit-and-stare IRIS data, enables direct comparisons with previous studies, and facilitates future comparative analyses. Furthermore, this analysis is meant as a direct continuation of \cite{Humphries_2024}.

Standard IRIS level 2 processing is applied to the dataset, incorporating dark current subtraction, geometric and orbital calibrations, and flat-field corrections. Over-saturated pixels are excluded from the filtering and detection method. The central slit and adjacent pixels remain unmodified and are not treated as missing data (see \paperi\ for further details). Additionally, sub-pixel image shifts caused by solar rotation are corrected using a Fourier Local Correlation Tracking method \citep{fw}.

\begin{figure*}[t]
    \centering
    \includegraphics[width=0.8\textwidth]{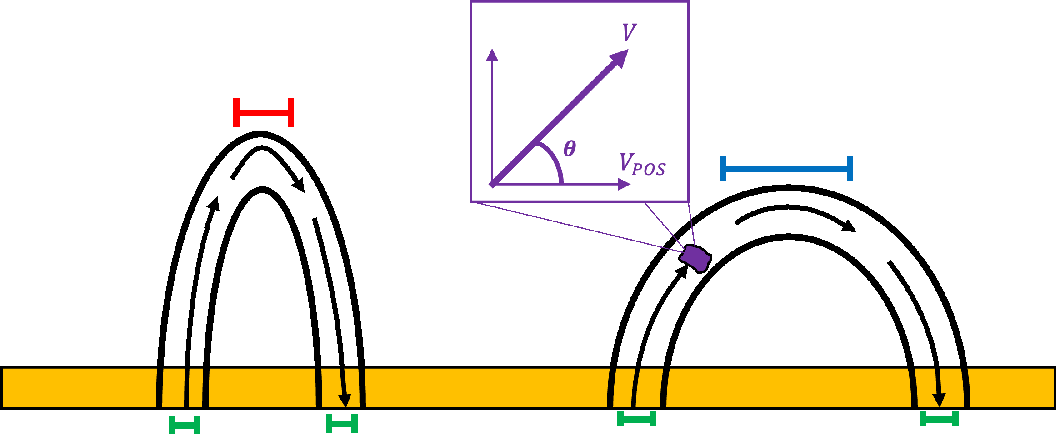}
    \caption{Proposed schematic diagram of magnetic loops representing those found in (left) Active and (right) Quiet domains. Both examples maximize the likelihood that BPs reach their maximum speeds during their lifetime's halfway point, at the apex of the arch. Green lines represent very short POS distances (and speeds), while red lines represent AQS POS distances (and speeds), and blue lines represent AQS (and the largest) POS distances (and speeds). The yellow layer represents the photosphere. The purple blob represents a BP as it traverses a loop, together with a vector diagram of its velocity.}
    \label{fig:schem_1}
\end{figure*}

\subsection{Detection Method}\label{sec:sub_detect_method}

The detection method and its application are detailed extensively in \cite{Humphries_2021_a}, \cite{Humphries_2021_b}, and \cite{Humphries_2024}. The same frequency limits for band-pass filtering (0.09 and 0.40 as a minimum and maximum, respectively) from previous studies are applied to this dataset. Candidate brightenings that are too small (less than 25 voxels) are discarded. For each detected BP, several attributes are recorded at each observation step throughout the BP’s lifetime. 
These step-by-step results are labeled "Of Each Step" (OES). The attributes include: total brightness, speed, travel distance, acceleration, and area. An additional attribute, intrinsic brightness, is also recorded, defined as the total brightness divided by both duration and mean area. 
All brightness values are background-subtracted, with the background intensity estimated as the median value of a surrounding boxed region. This box extends a few pixels beyond the BP’s minimum and maximum spatial coordinates in both x- and y-directions. The background intensity is determined using pixel values within this box but outside the BP itself. 
An \nfrag\ parameter is also recorded, which describes the largest number of isolated regions of an event during its lifetime (see \cite{Humphries_2021_a} and \cite{Humphries_2021_b} for details).

\begin{sidewaysfigure}[htbp]
    \centering
    \begin{minipage}{0.19\textwidth} 
        \centering
        (a)\includegraphics[width=\linewidth]{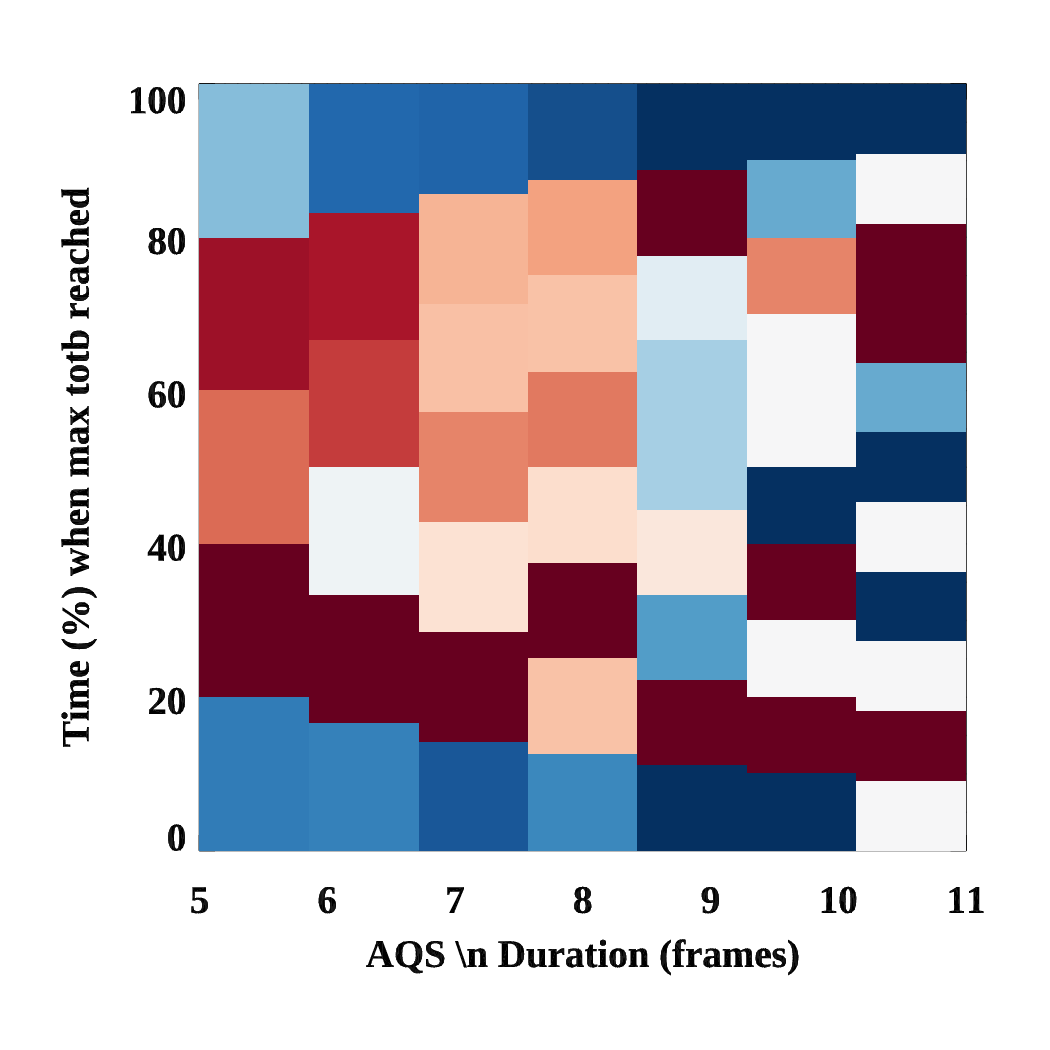}
    \end{minipage}
    \begin{minipage}{0.19\textwidth}
        \centering
        (b)\includegraphics[width=\linewidth]{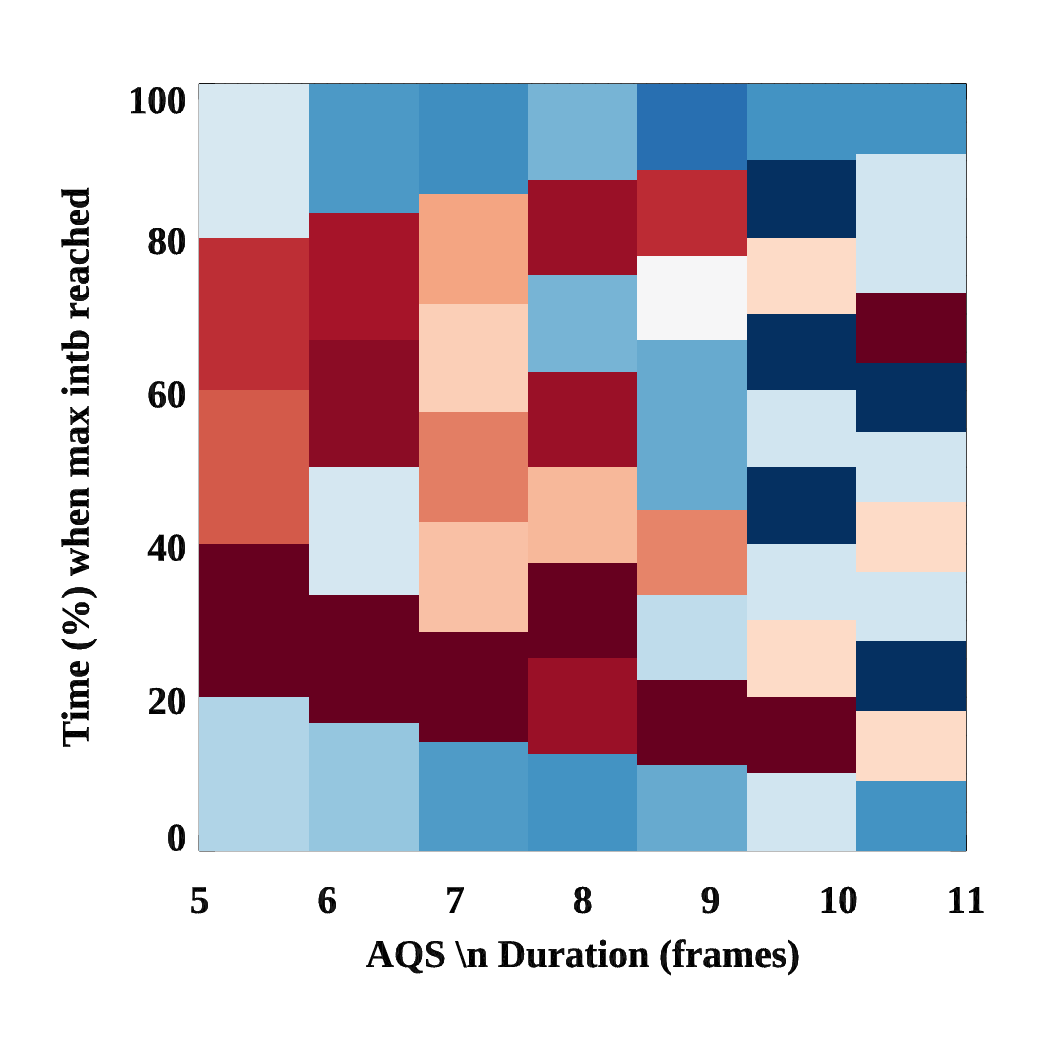}
    \end{minipage}
    \begin{minipage}{0.19\textwidth}
        \centering
        (c)\includegraphics[width=\linewidth]{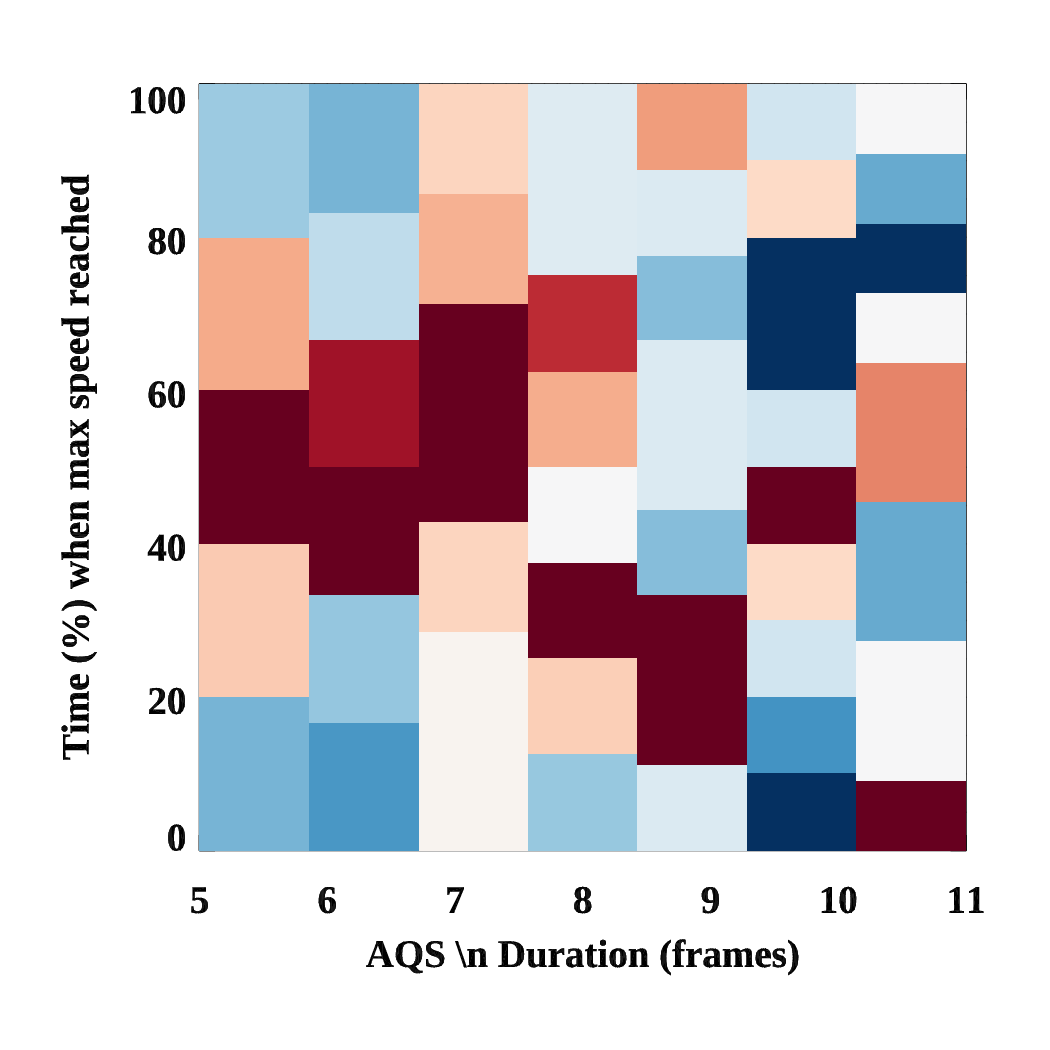}
    \end{minipage}
    \begin{minipage}{0.19\textwidth}
        \centering
        (d)\includegraphics[width=\linewidth]{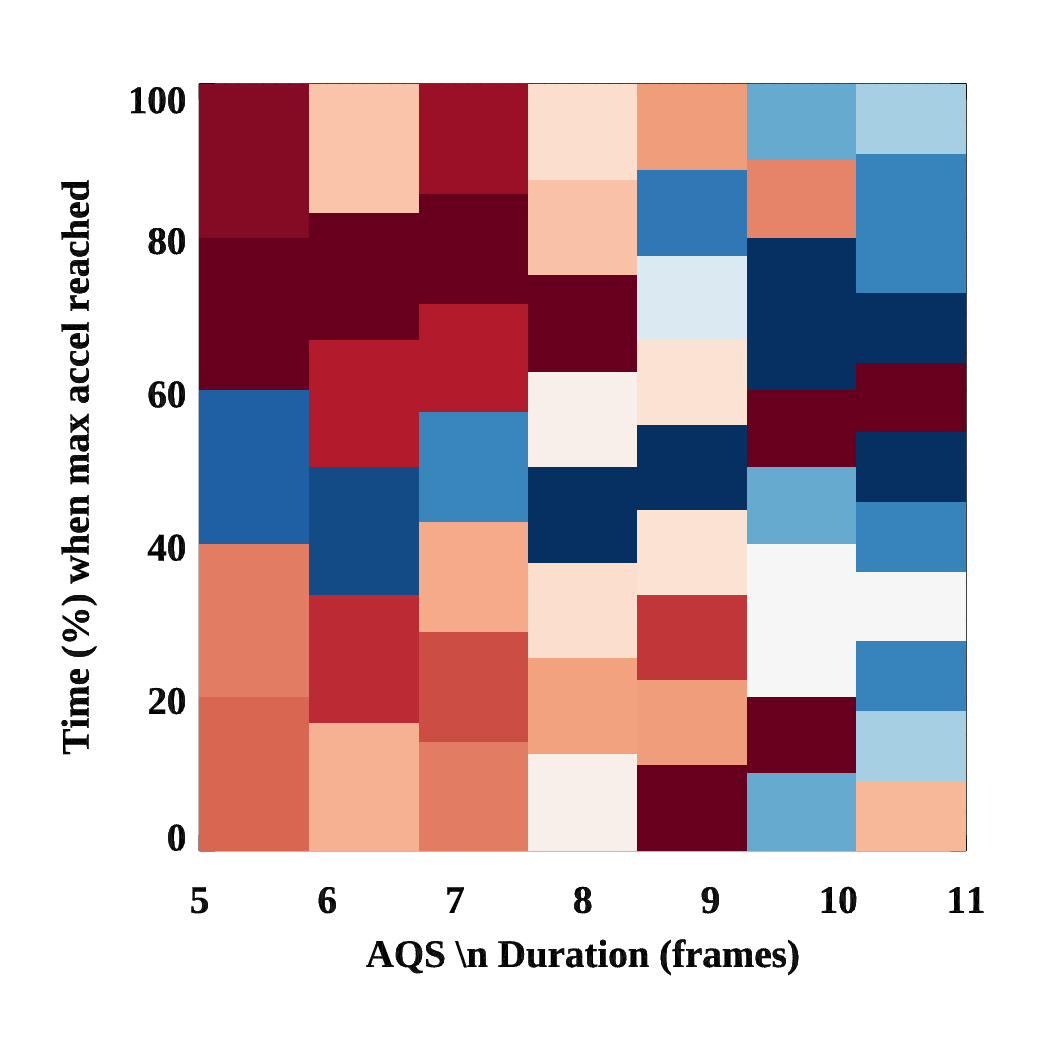}
    \end{minipage}
    \begin{minipage}{0.19\textwidth}
        \centering
        (e)\includegraphics[width=\linewidth]{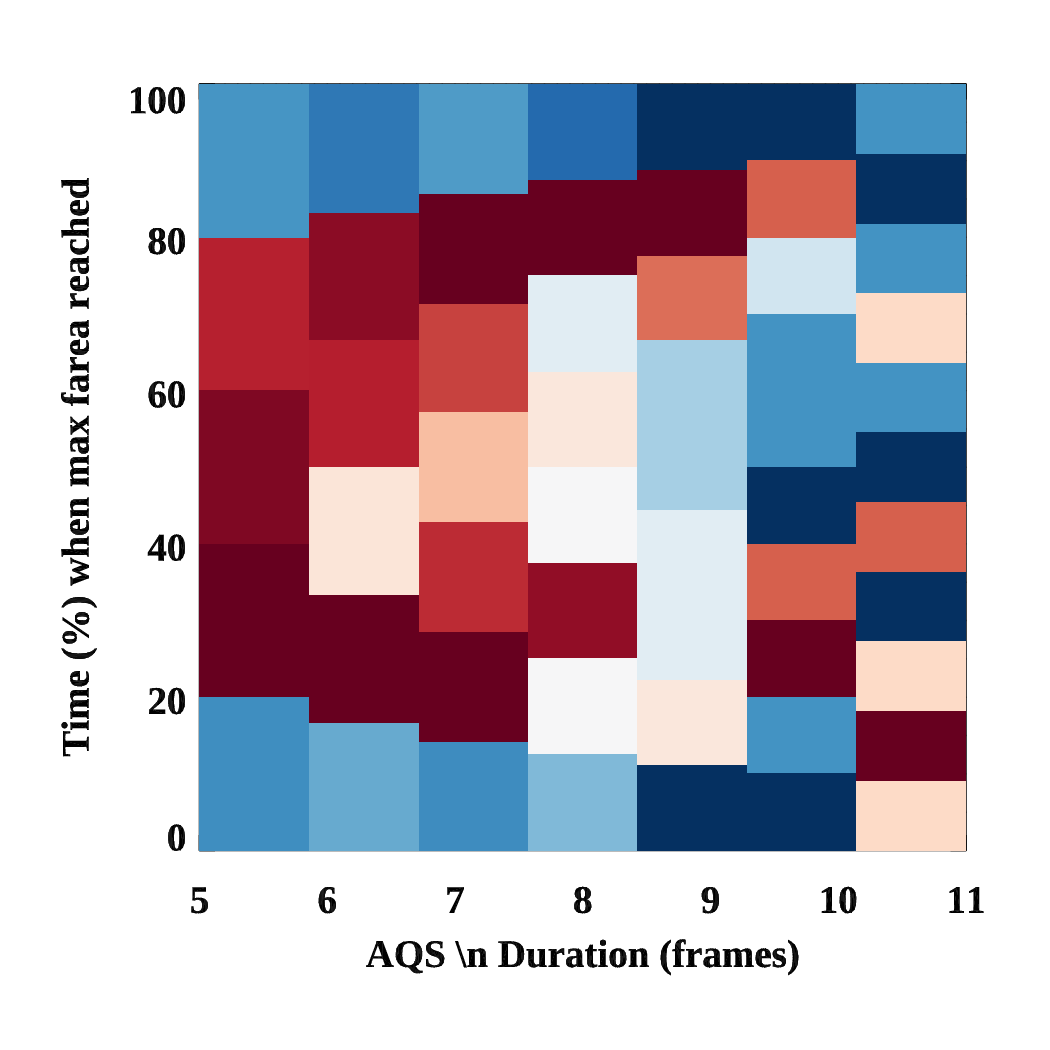}
    \end{minipage}

    \begin{minipage}{0.19\textwidth} 
        \centering
        (f)\includegraphics[width=\linewidth]{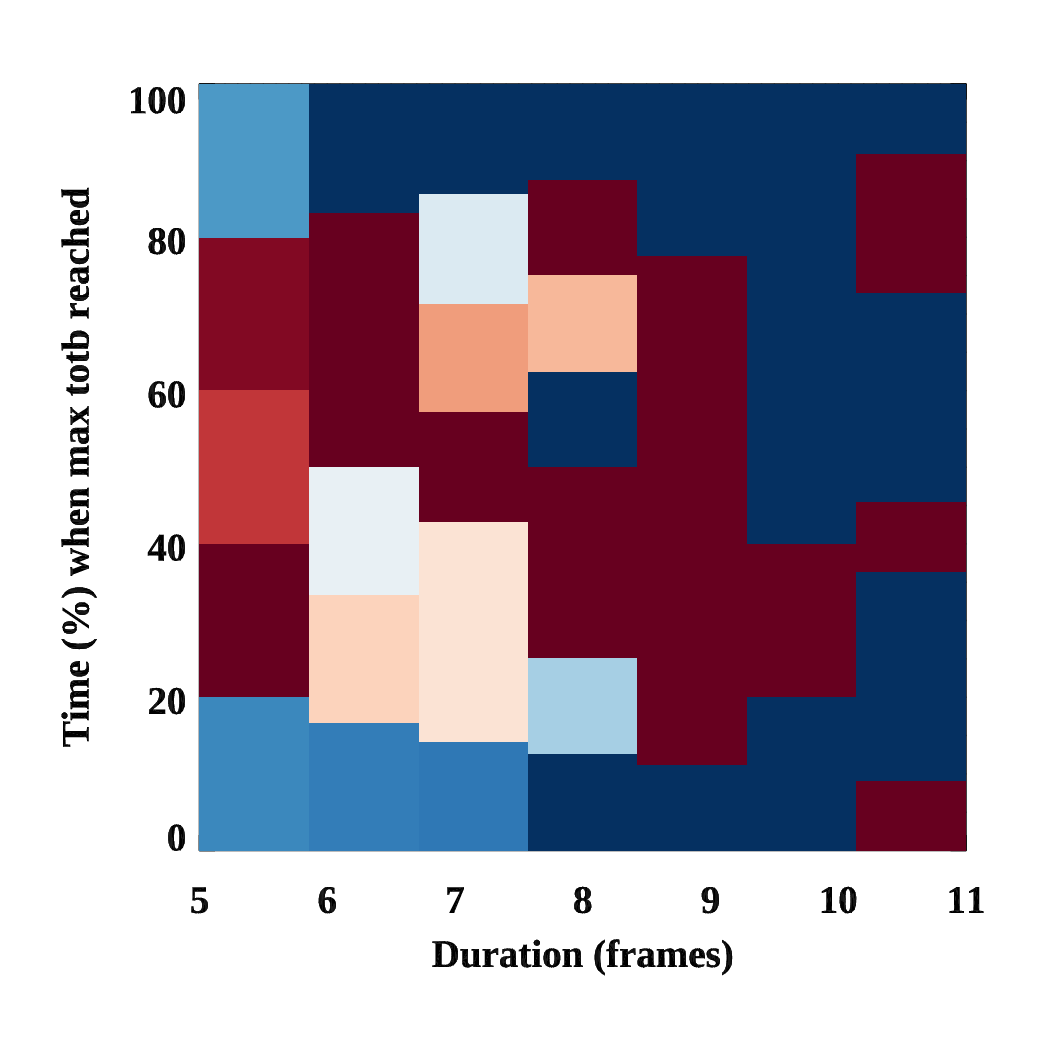}
    \end{minipage}
    \begin{minipage}{0.19\textwidth}
        \centering
        (g)\includegraphics[width=\linewidth]{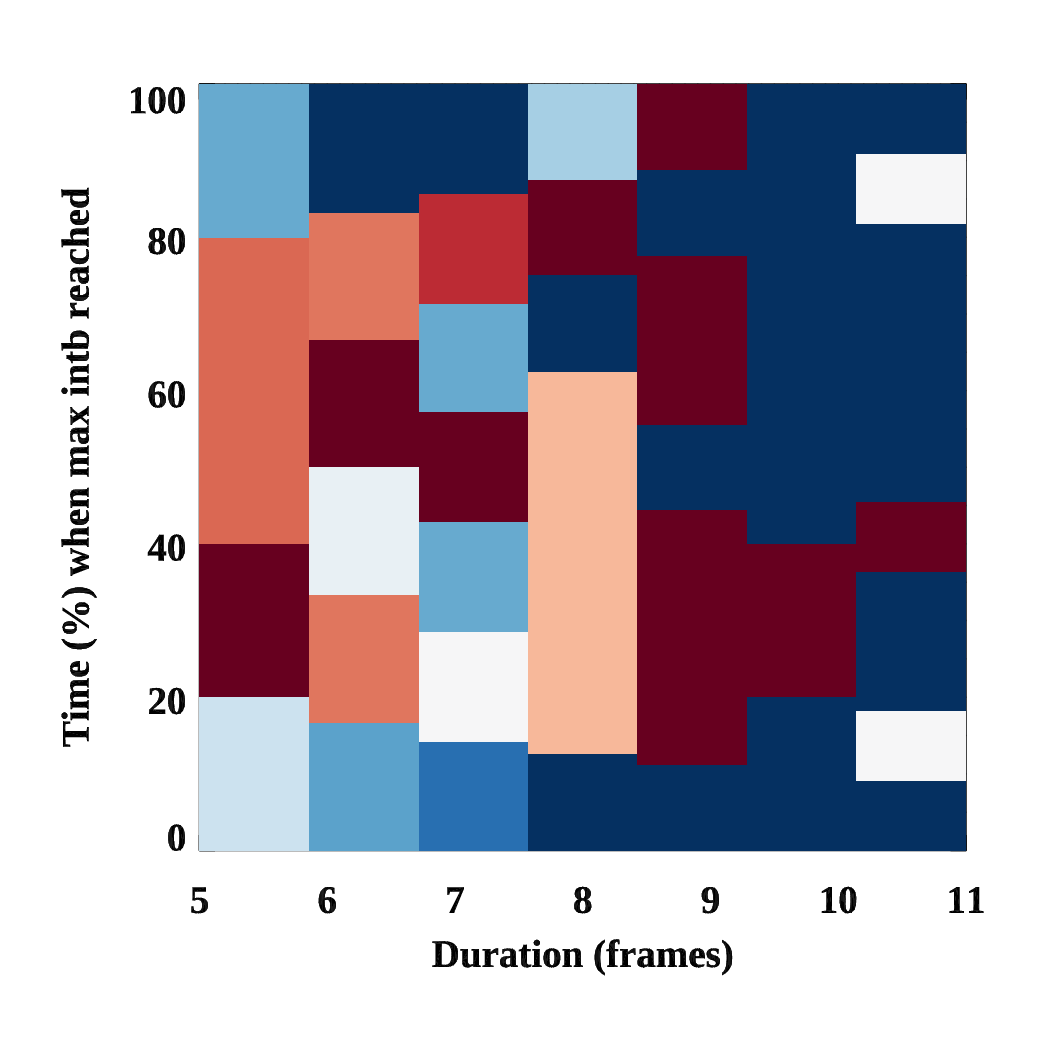}
    \end{minipage}
    \begin{minipage}{0.19\textwidth}
        \centering
        (h)\includegraphics[width=\linewidth]{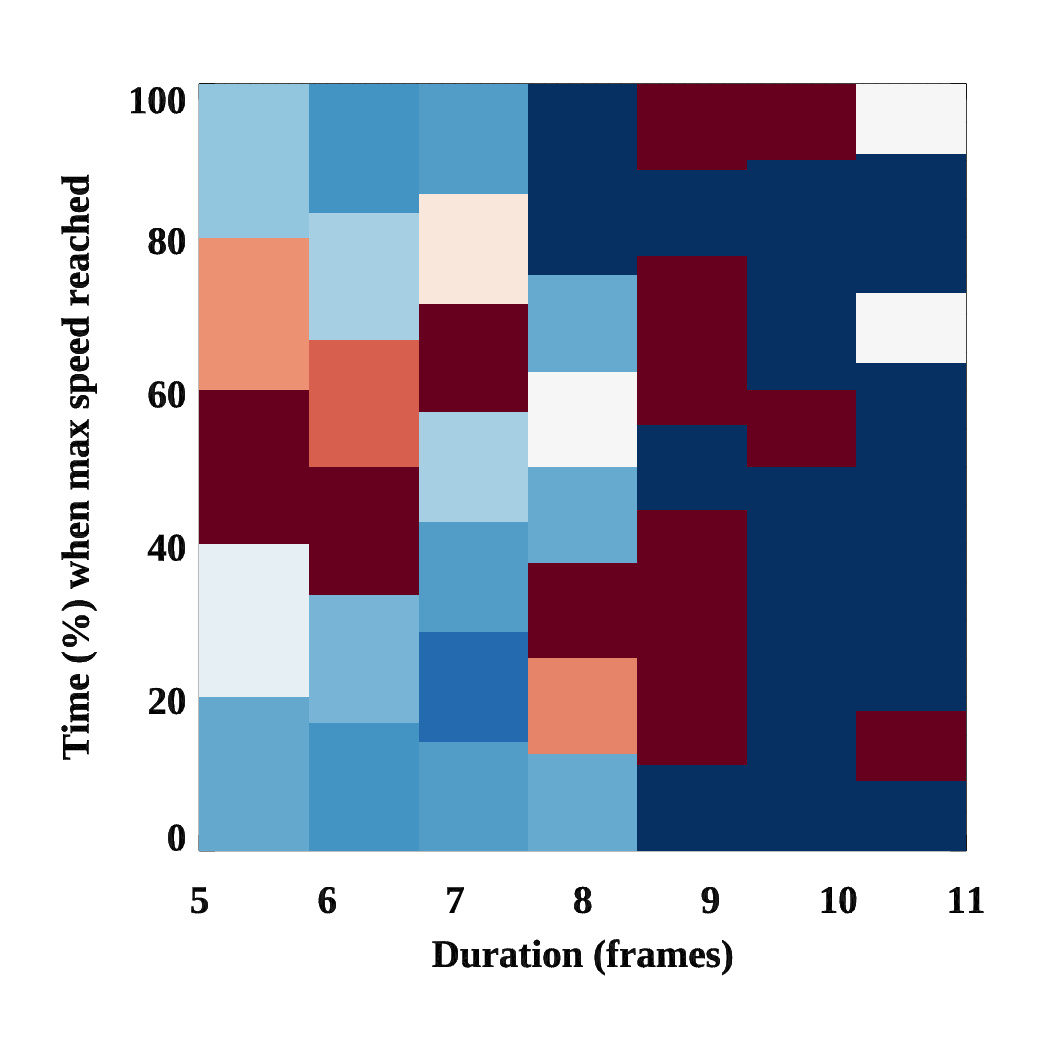}
    \end{minipage}
    \begin{minipage}{0.19\textwidth}
        \centering
        (i)\includegraphics[width=\linewidth]{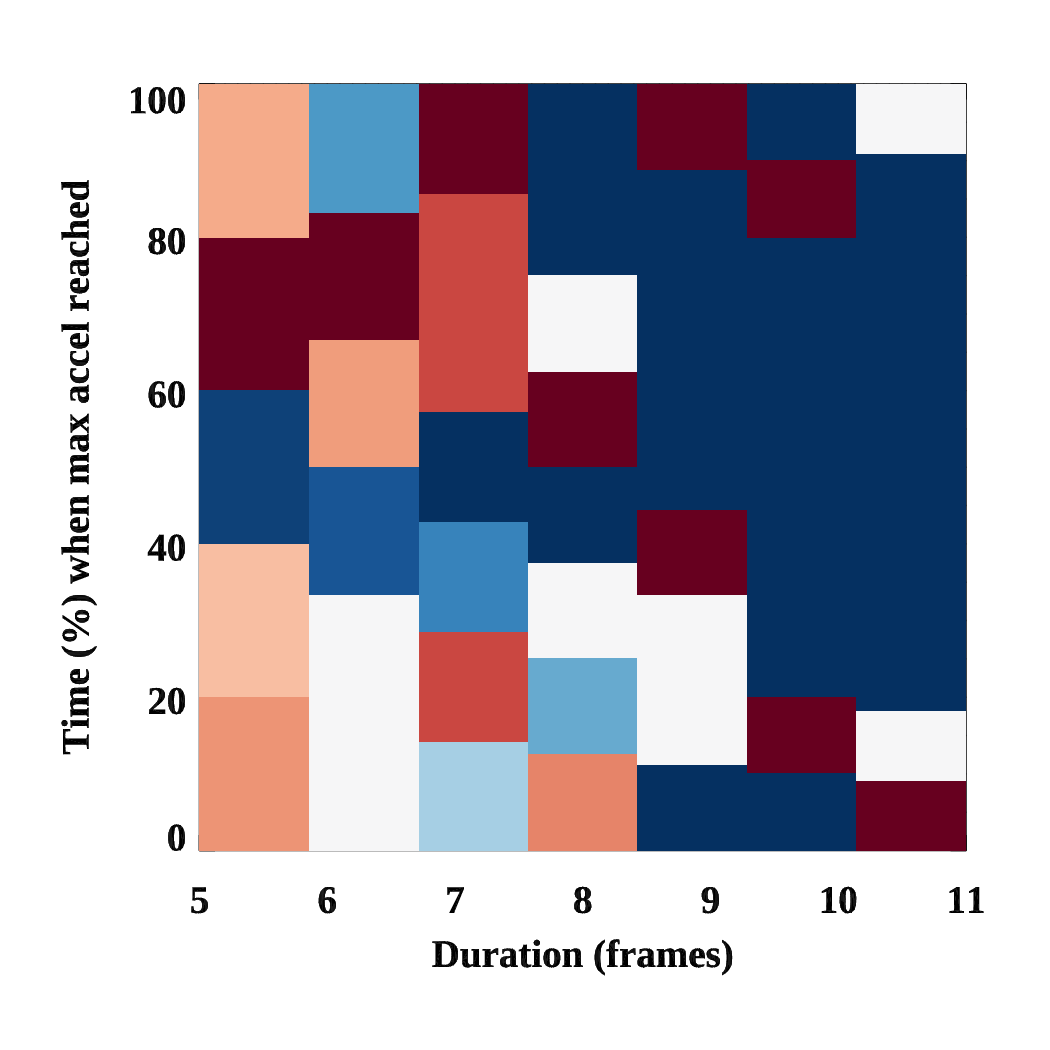}
    \end{minipage}
    \begin{minipage}{0.19\textwidth}
        \centering
        (j)\includegraphics[width=\linewidth]{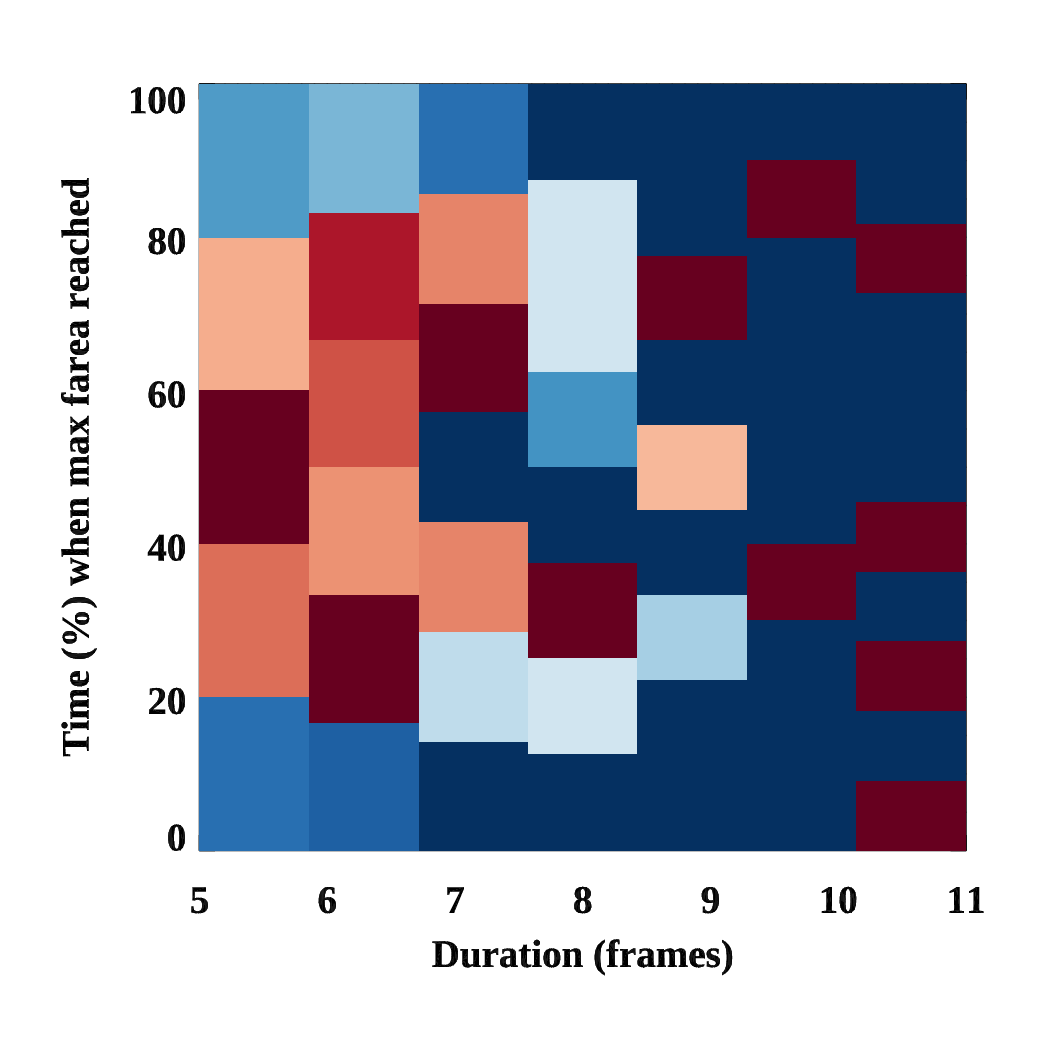}
    \end{minipage}

    \vspace{0.5cm} 
    \begin{minipage}{0.19\textwidth} 
        \centering
        (k)\includegraphics[width=\linewidth]{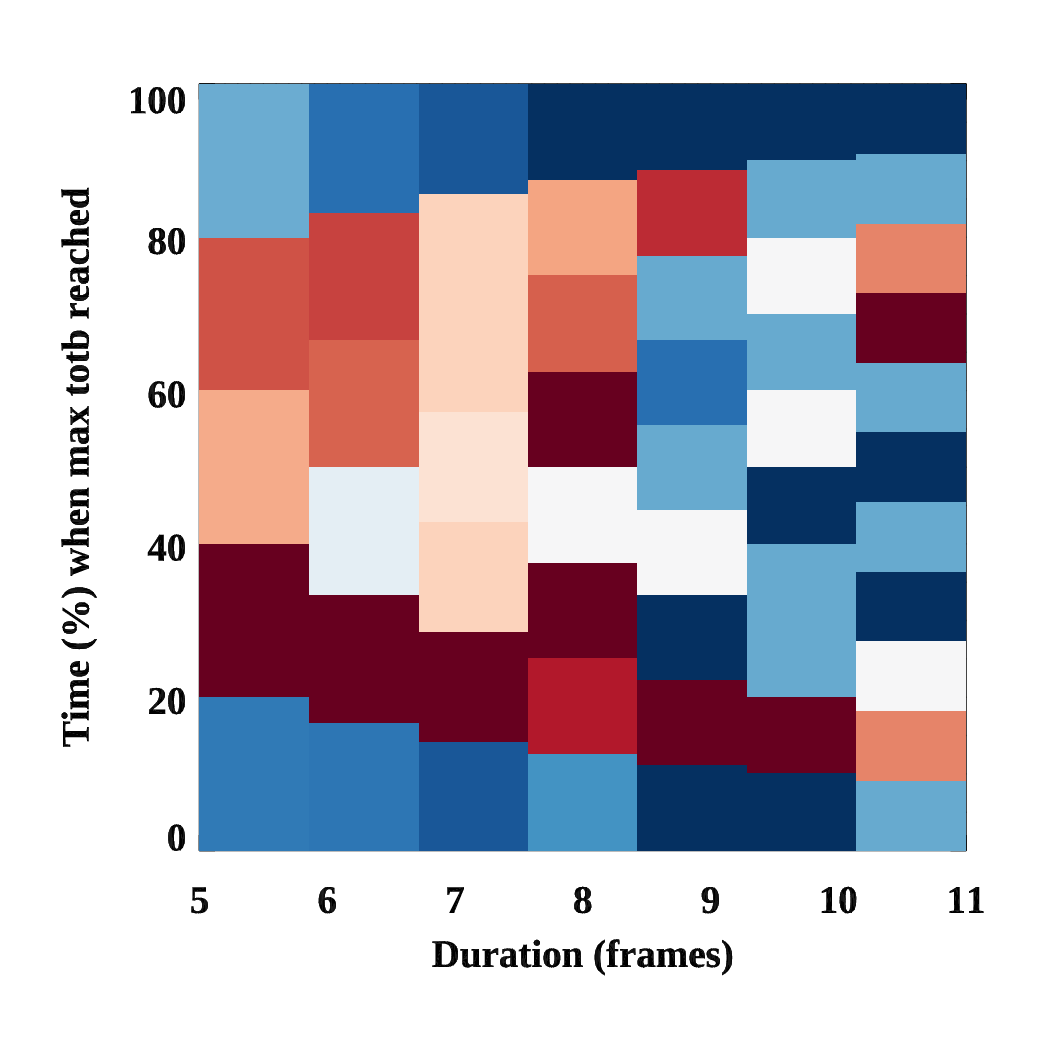} 
    \end{minipage}
    \begin{minipage}{0.19\textwidth}
        \centering
        (l)\includegraphics[width=\linewidth]{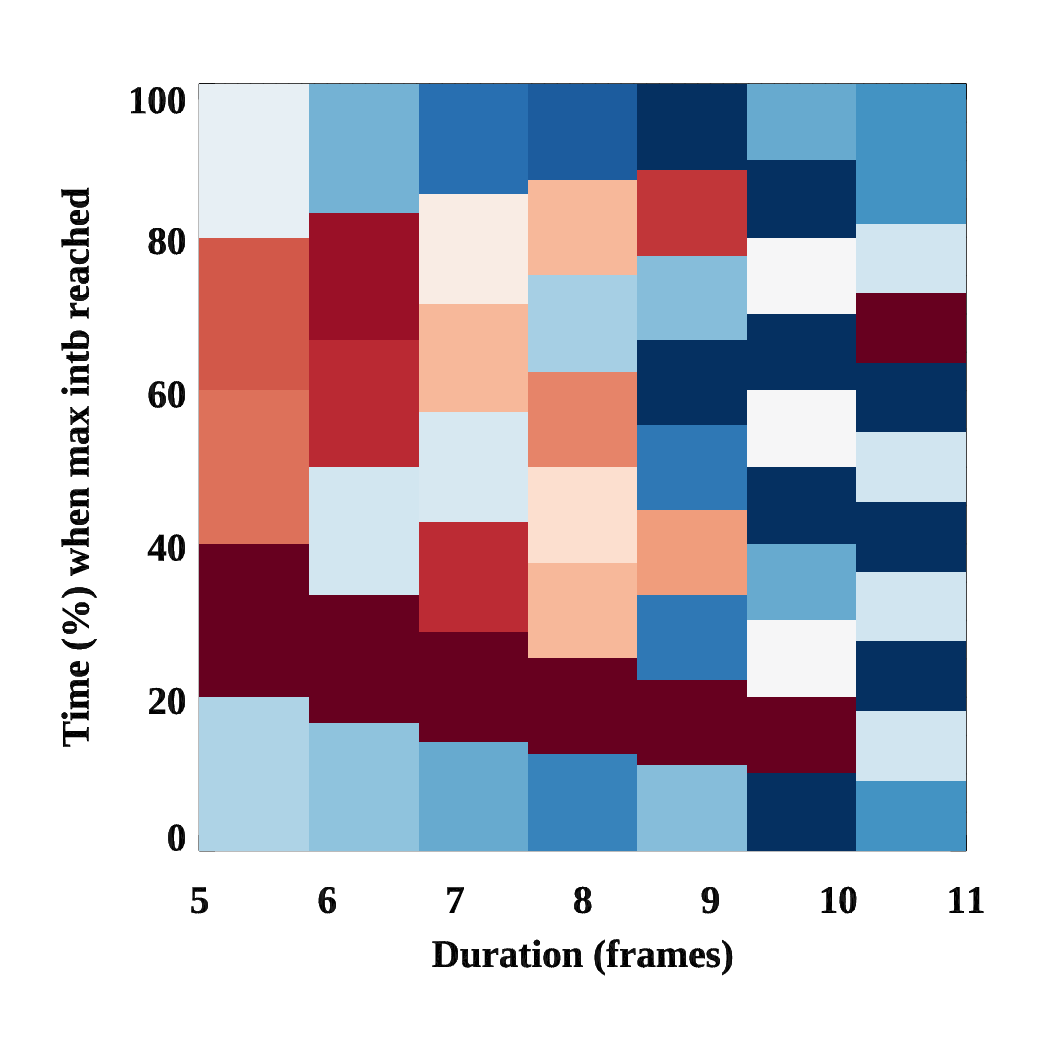}
    \end{minipage}
    \begin{minipage}{0.19\textwidth}
        \centering
        (m)\includegraphics[width=\linewidth]{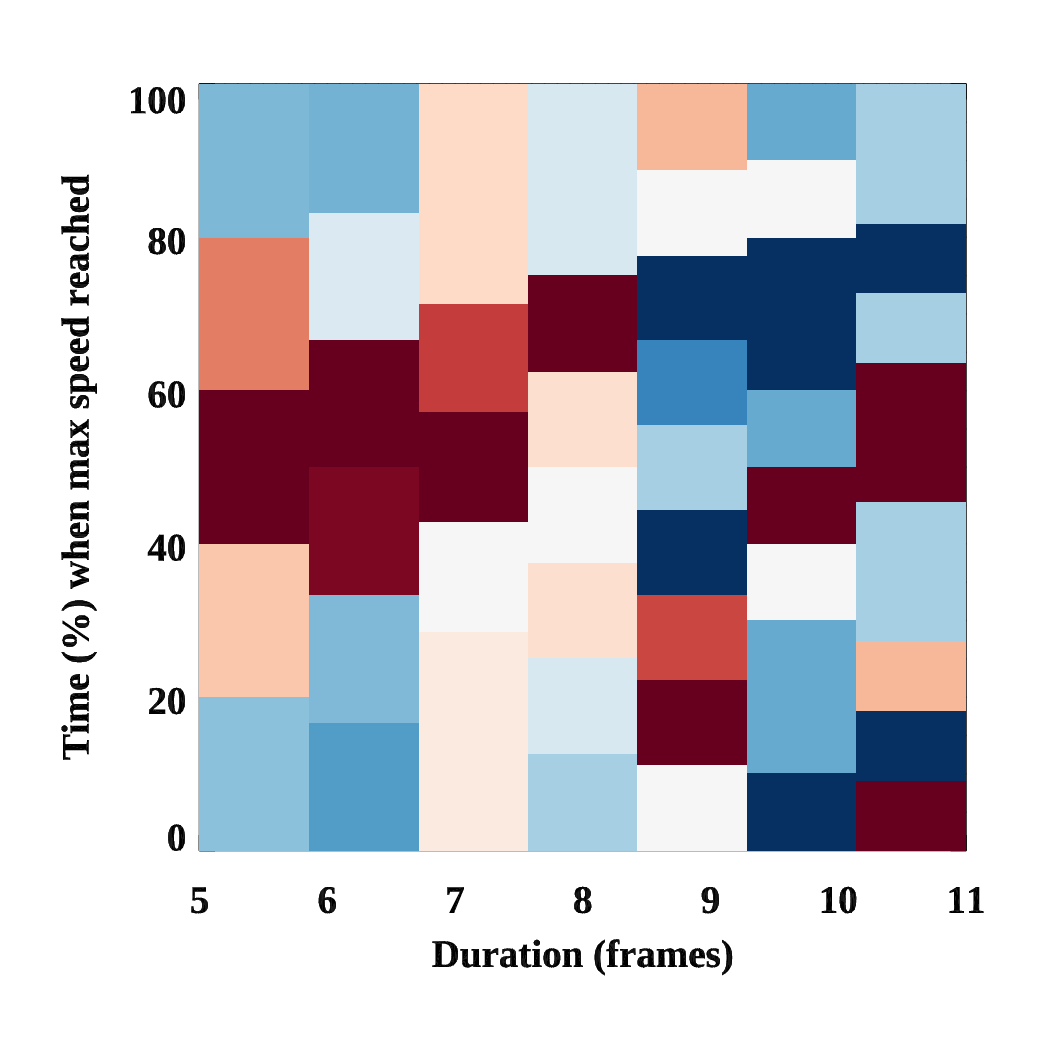}
    \end{minipage}
    \begin{minipage}{0.19\textwidth}
        \centering
        (n)\includegraphics[width=\linewidth]{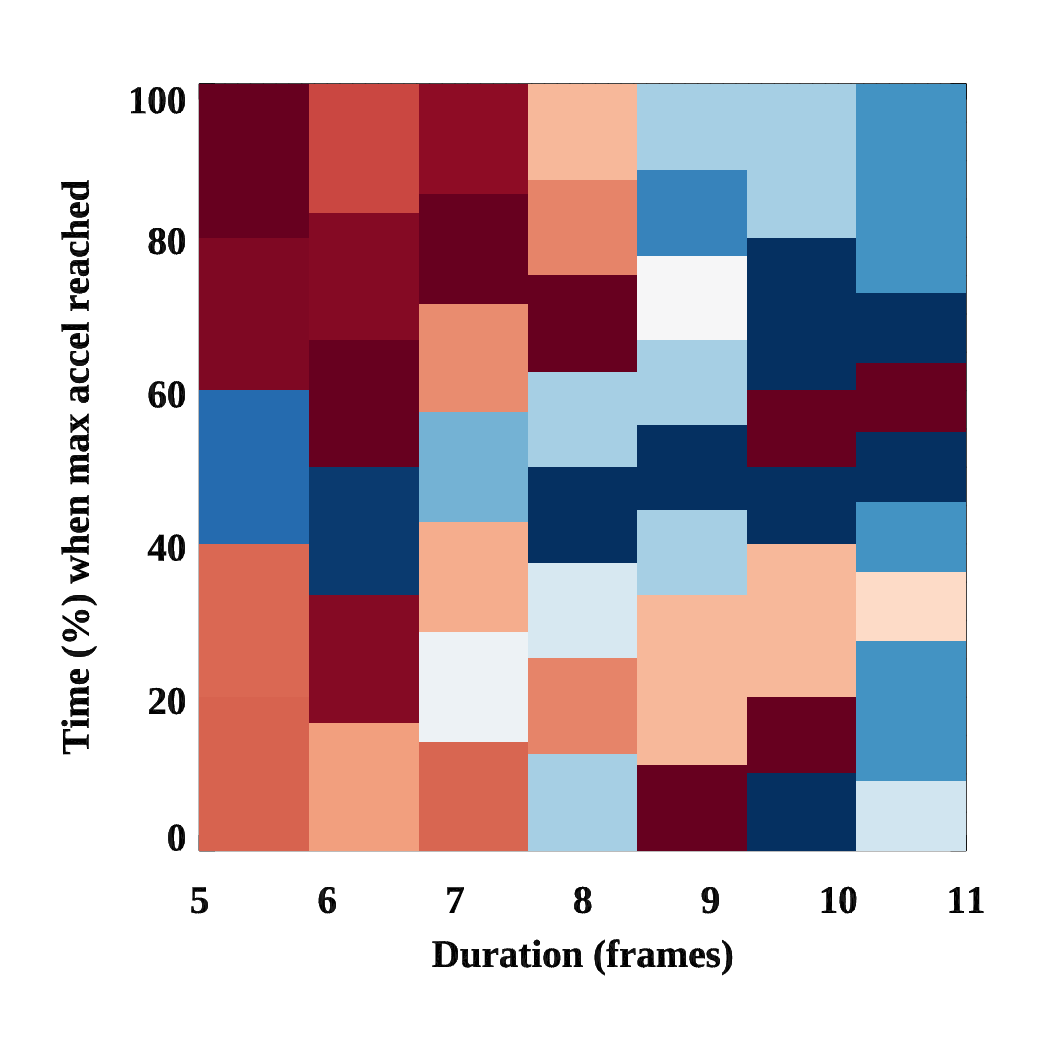}
    \end{minipage}
    \begin{minipage}{0.19\textwidth}
        \centering
        (o)\includegraphics[width=\linewidth]{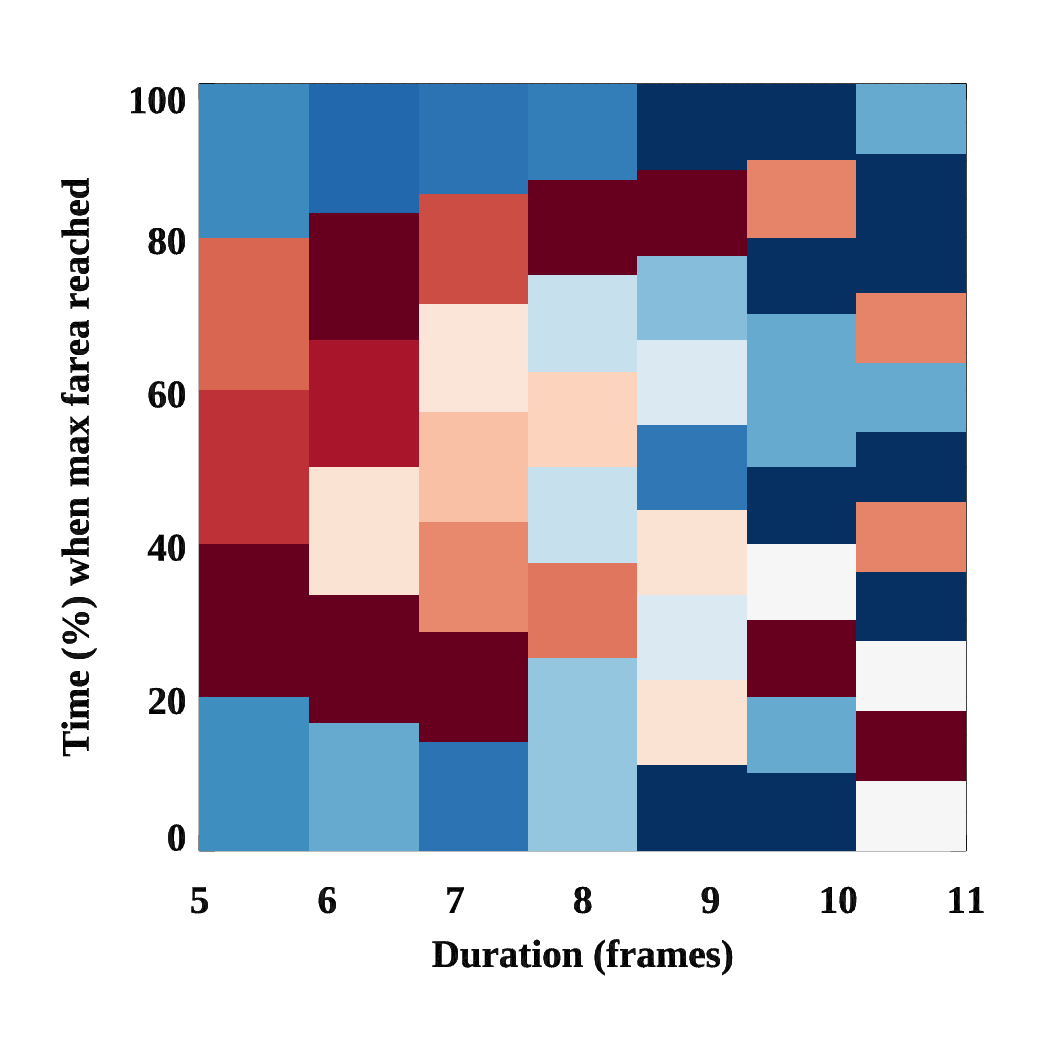}
    \end{minipage}

    \caption{2D histogram plots of the temporal position (\%) at which (from left to right), total brightness, intrinsic brightness, POS speed, acceleration, and area reach their maximum values vs BP duration (in frames). Top, middle and bottom rows represent results for all BPs, AQS BPs and TQS BPs, respectively.}
    \label{fig:attribute_vs_dur}
\end{sidewaysfigure}

\subsection{Analysis Method}\label{sec:sub_analysis_method}

We select BPs with \nfrag$=1$ and a minimum duration of 5 frames. The restriction to \nfrag$=1$ is consistent with previous studies for simplicity, though it differs from \cite{Humphries_2024}, which also includes \nfrag$=2$. This adjustment is to avoid any additional complications in interpreting results. The 5-frame minimum duration is imposed to mitigate potential biases from very short-lived (SL) BPs in plotting. This approach modifies the 3-frame lower limit used in \cite{Humphries_2024}, which corresponds to a change from a minimum duration of 85 seconds to that of $\sim135$ seconds. Implementing these restrictions reduces the total number of detected BPs from $\sim14,600$ to $\sim1300$. To standardize comparisons, the time series of each recorded attribute, such as total brightness, is expanded over 100 bins using quadratic interpolation.

We implement the same ``Active" Quiet-Sun (AQS) and ``True" Quiet-Sun (TQS) domain differentiation as that of \cite{Humphries_2024} by determining the mean value of the time-averaged 1400 \AA\ data. AQS domains are defined as regions with brightness $\ge1.4$ times the mean brightness across the whole region, and lie within the red countours in figure \ref{fig:FOV_w_contours}.  TQS are defined as $\le1.2$ times the mean and lie within the blue countours in figure \ref{fig:FOV_w_contours}. If the mean spatial position of BPs are detected within either of these domains, then they are recorded as either AQS BPs or TQS BPs accordingly. Domains that fall between these values (as well as any BPs detected therein) are ignored (see \cite{Humphries_2024} for further details).

\begin{figure}[t]
    \centering
    \includegraphics[width=0.45\linewidth]{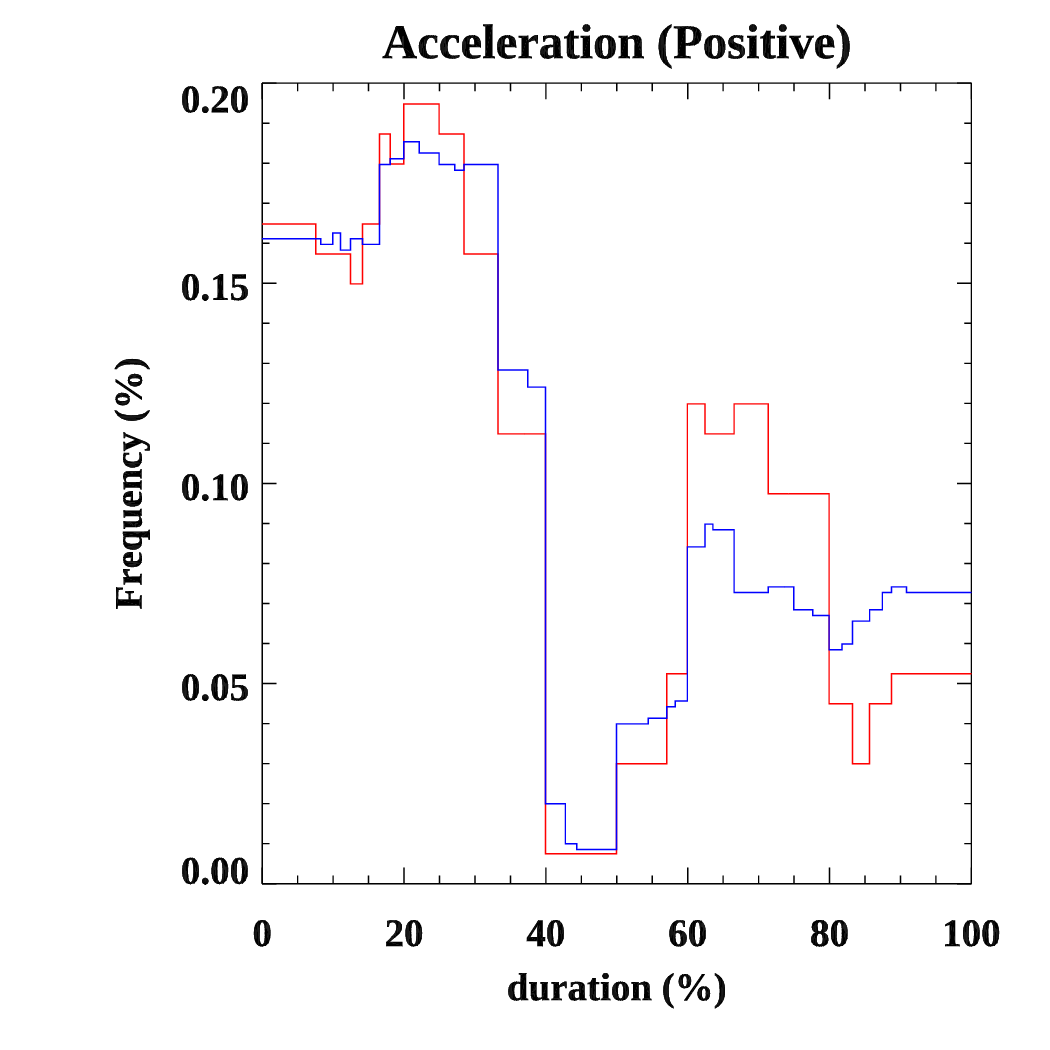}
    \includegraphics[width=0.45\linewidth]{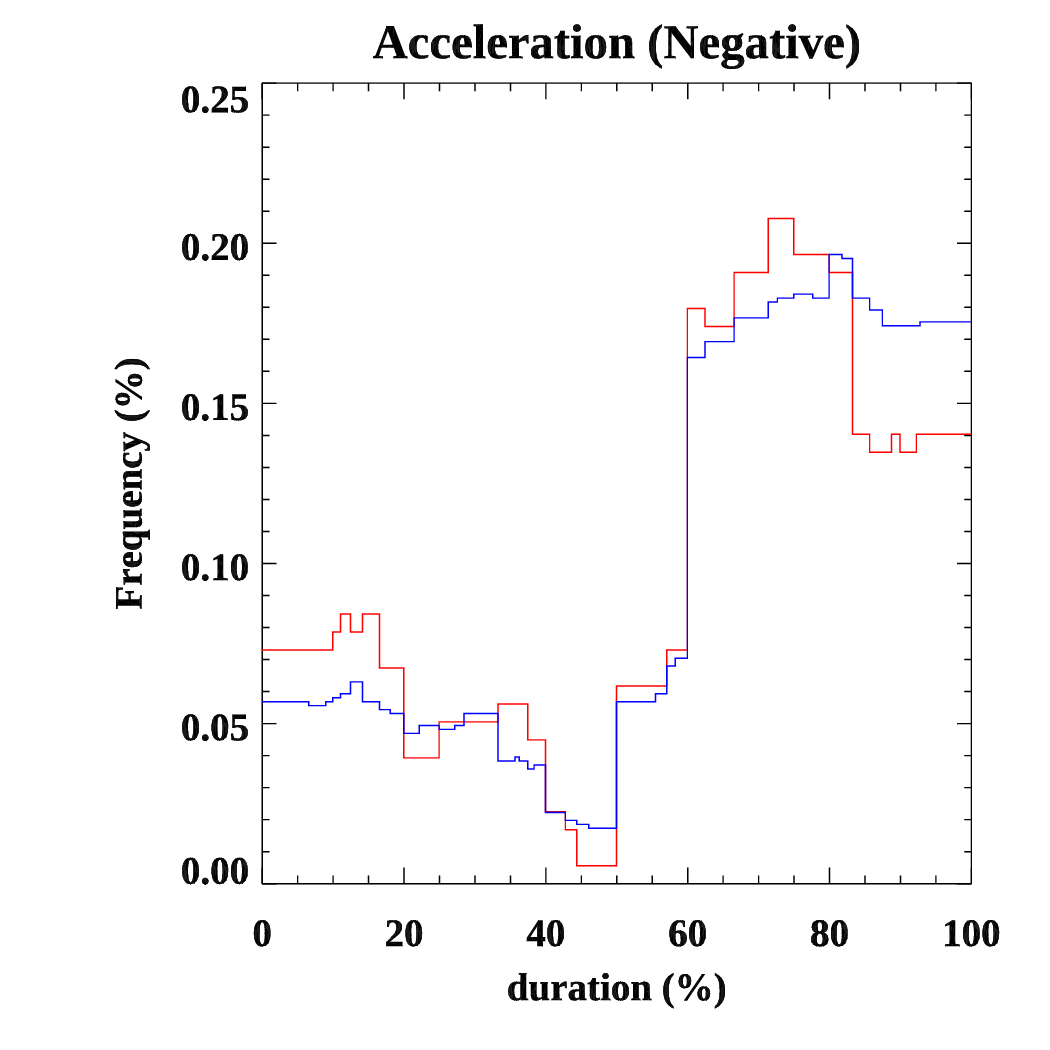}
    \caption{Histogram temporal comparisons of when (a) positive acceleration and (b) negative accelertation reach their maximum values. These results correspond to separating the absolute acceleration values in figure \ref{fig:max_temp_pos}d. Red and blue lines represent AQS and TQS results, respectively.}
    \label{fig:pos_neg_1D}
\end{figure}

\section{Results \& Discussion} \label{sec:results}

\subsection{Histogram results}

Figure \ref{fig:max_temp_pos} presents histograms showing the temporal position (expressed as a percentile) at which BPs reach their maximum values for various parameters: (a) total brightness, (b) intrinsic brightness, (c) POS speed, (d) POS absolute acceleration, and (e) area. The production of these 1D histograms - whereby the number of time bins, and the greater number of BPs with shorter durations, can introduce biases without proper consideration - is described in the appendix. Red and blue distributions correspond to BPs in 'Active' and 'Quiet' domains, respectively. We also analyse travel distances; however, the resulting plots exhibit the same shape as those of speed, since speed is derived from travel distance, and are therefore not shown here.

It is possible that results at the extreme ends of these distributions (0th and 100th percentiles) are afflicted by some detection biases. For instance, in the case of total brightness, hard lower- and upper-bound cut-offs exist depending on the intensity of individual pixels within the dataset. 

Figures \ref{fig:max_temp_pos}(a), (b), and (e) indicate some preference for when TQS BPs reach their maximum brightness or size - primarily near the 30th and 70th percentiles of their lifetimes. The likelihood of AQS BPs to reach these maximums is fairly evenly distributed between these percentile points. BPs are also generally less likely to achieve their maximum values during the first and final fifth of their lifetime.

Speed and acceleration behave differently from other attributes. Speed exhibits a clear peak at the 50th percentile for both AQS and TQS BPs with a gradual drop on either side, suggesting that the majority of BPs reach their maximum POS speed around the midpoint of their lifetime. Acceleration, by contrast, shows significant peaks within the first and final 40\% of a BP’s lifetime, on either side of the maximum speed peak. The second acceleration peak is slightly asymmetric with a smaller peak near the 50th percentile, indicating that some BPs reach their maximum acceleration near the midway point of their lifetime. Another assymmetry exists with AQS BP acceleration within the final 40\% of their lfietime, whereby the likelihood peaks near 70\% before dropping off closer to the end of their lifetimes.

Figure \ref{fig:attribute_vs_dur} presents a 2D histogram illustrating the relationship between BP duration and the temporal percentile at which various BP attributes reach their maximum values. These attributes include: (first column) total brightness, (second column) intrinsic brightness, (third column) POS speed, (fourth column) POS absolute acceleration, and (fifth column) area. This figure is generated in a similar fashion to that of figure \ref{fig:max_temp_pos} and acts as its complement by offering an alternative perspective on when, during a BP’s lifetime, each property is most likely to peak.
BPs with durations exceeding 11 frames are re-binned into the 11th frame due to increasing noise and scattered distributions in longer-lived examples.
The top row represents the results from all BPs. Total brightness typically peaks around 20\% into a BP’s lifetime, irrespective of the BP’s total duration. A secondary peak is observed near the 80th percentile, consistent with patterns noted in Figure \ref{fig:max_temp_pos}(a). Some dissimilar behaviour can be seen between short-lived (SL) and long-lived (LL) BPs, whereby SL BPs can reach their peak total brightness at almost any stage of their lifetime, whereas LL BPs are more likely to peak early.
Intrinsic brightness follows a similar trend and SL/LL variability, with the highest probability of reaching maximum values near the 20th percentile.
POS speed, as shown in figure \ref{fig:attribute_vs_dur}(c), most commonly reaches its maximum around the midpoint (50\%) of a BP’s lifetime. This central peak diminishes in probability each side of the peak.
This mid-lifetime speed peak is more distinct in SL BPs. As duration increases, the timing of maximum POS speed becomes more dispersed, though LL BPs show a slight tendency to peak earlier.
Early-lifetime acceleration peaks appear relatively consistent across BPs of all durations, while late-lifetime acceleration peaks are more common in SL BPs.
Area tends to peak around both the 20th and 80th percentiles, regardless of BP lifetime. SL BPs may reach their maximum area at almost any point in their evolution, excluding the very start or end.
For longer-lived BPs, a stronger preference for area maxima near the 80th percentile is observed. This could reflect either the gradual diffusive nature of BPs or magnetic loop expansion over time.
While both total brightness and area often peak around similar times, total brightness more frequently peaks near 20\%, whereas area more frequently peaks near 80\%, reinforcing the interpretation that BPs become more diffuse as they evolve.

Magnetic reconnection remains the most likely physical driver, regardless of whether BPs trace direct plasma motion or result from wavefront-induced heating.
POS acceleration typically peaks near the 30th and 70th percentiles, with occasional peaks around the midpoint, expressing the same trends observed in Figure \ref{fig:max_temp_pos}(e).
Reconnection is likely to occur in the photosphere, with the observed delay of ~85–160 seconds between reconnection and peak brightness (total and intrinsic) suggesting either:
\begin{itemize}
    \item Delayed heating as plasma propagates upward, or
    \item A lag between reconnection and the detectable thermal impact of the associated shockwave in the chromosphere
\end{itemize}

Figure \ref{fig:attribute_vs_dur}f-o further breaks down the results of figure \ref{fig:attribute_vs_dur}a-e into (middle row, f-j) AQS and (bottom row, k-o) TQS components. 

\begin{figure}[t]
    \centering
    (a)\includegraphics[width=0.4\linewidth]{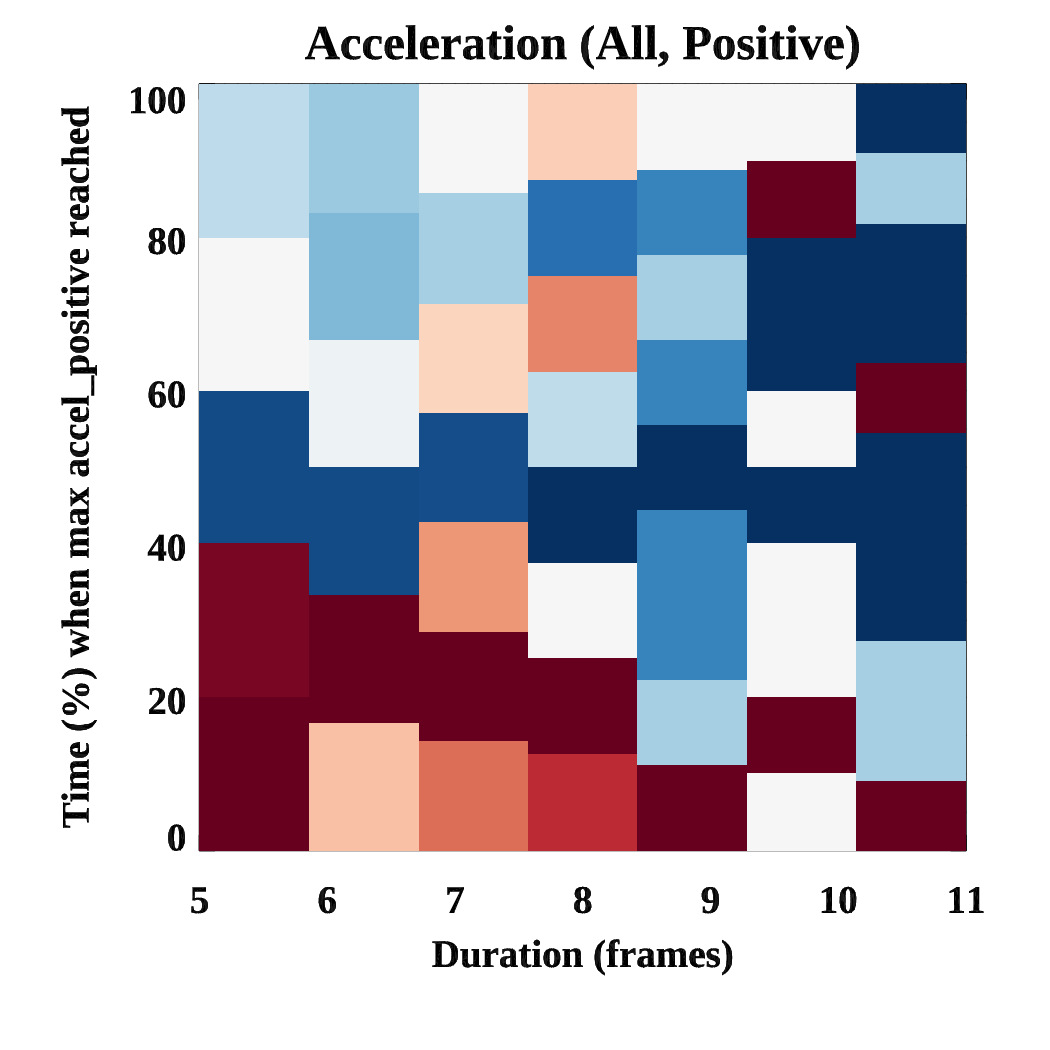}
    (b)\includegraphics[width=0.4\linewidth]{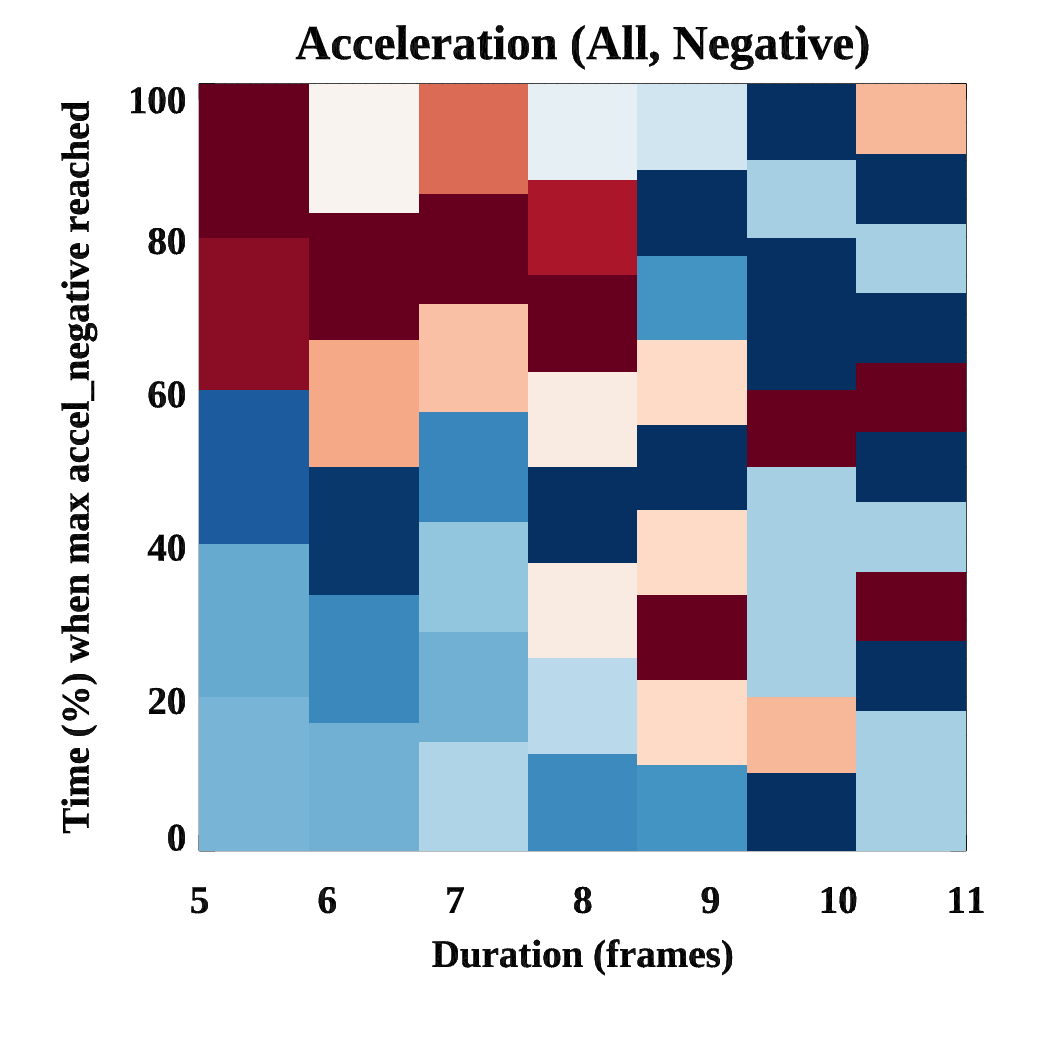}
    (c)\includegraphics[width=0.4\linewidth]{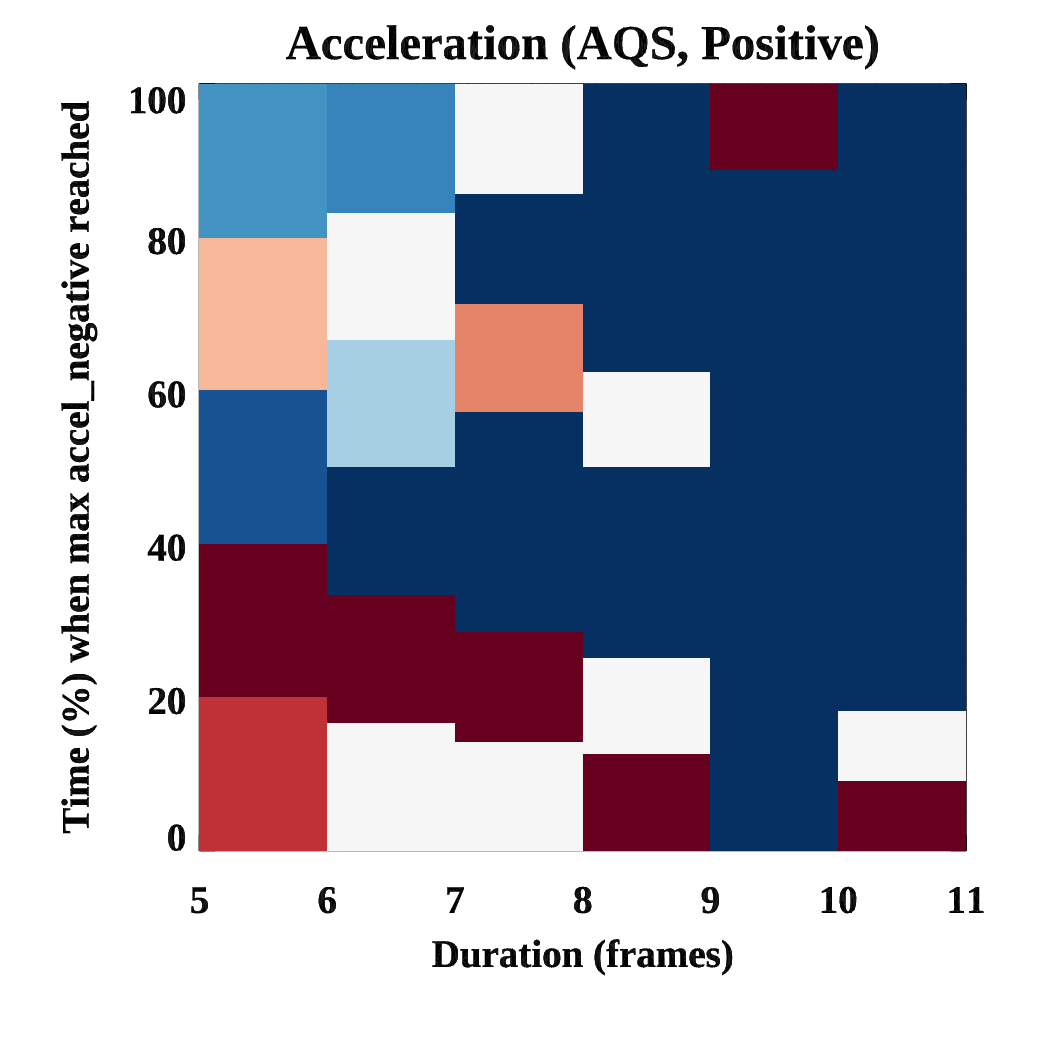}
    (d)\includegraphics[width=0.4\linewidth]{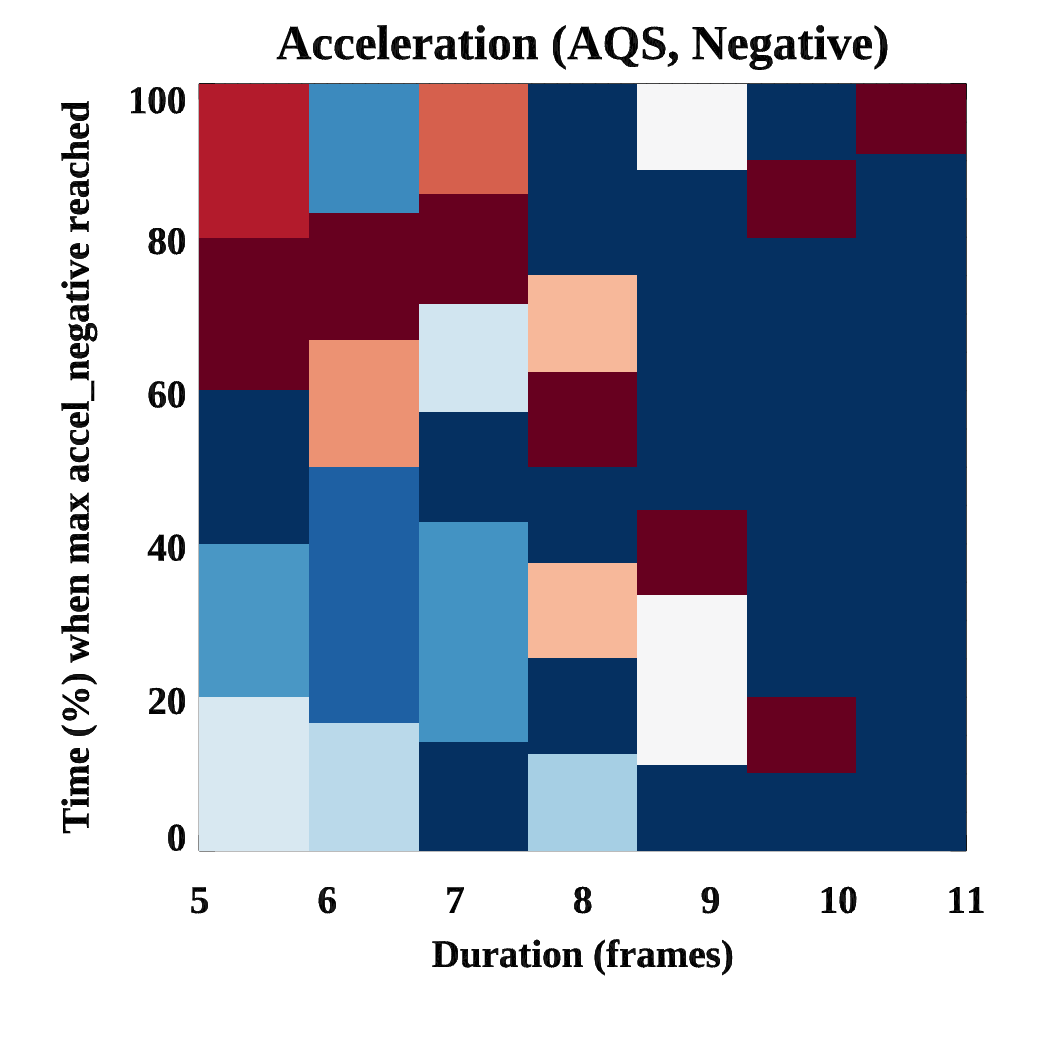}
    (e)\includegraphics[width=0.4\linewidth]{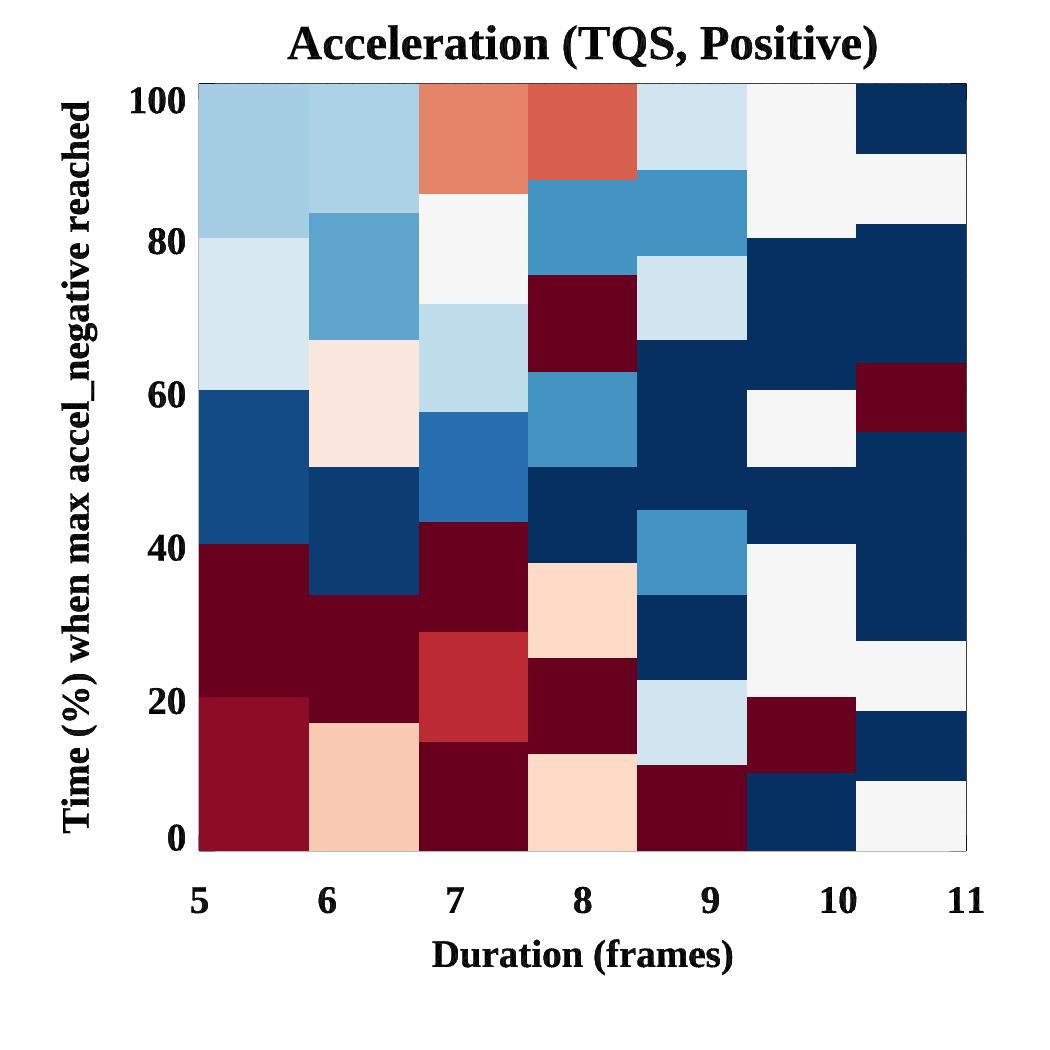}
    (f)\includegraphics[width=0.4\linewidth]{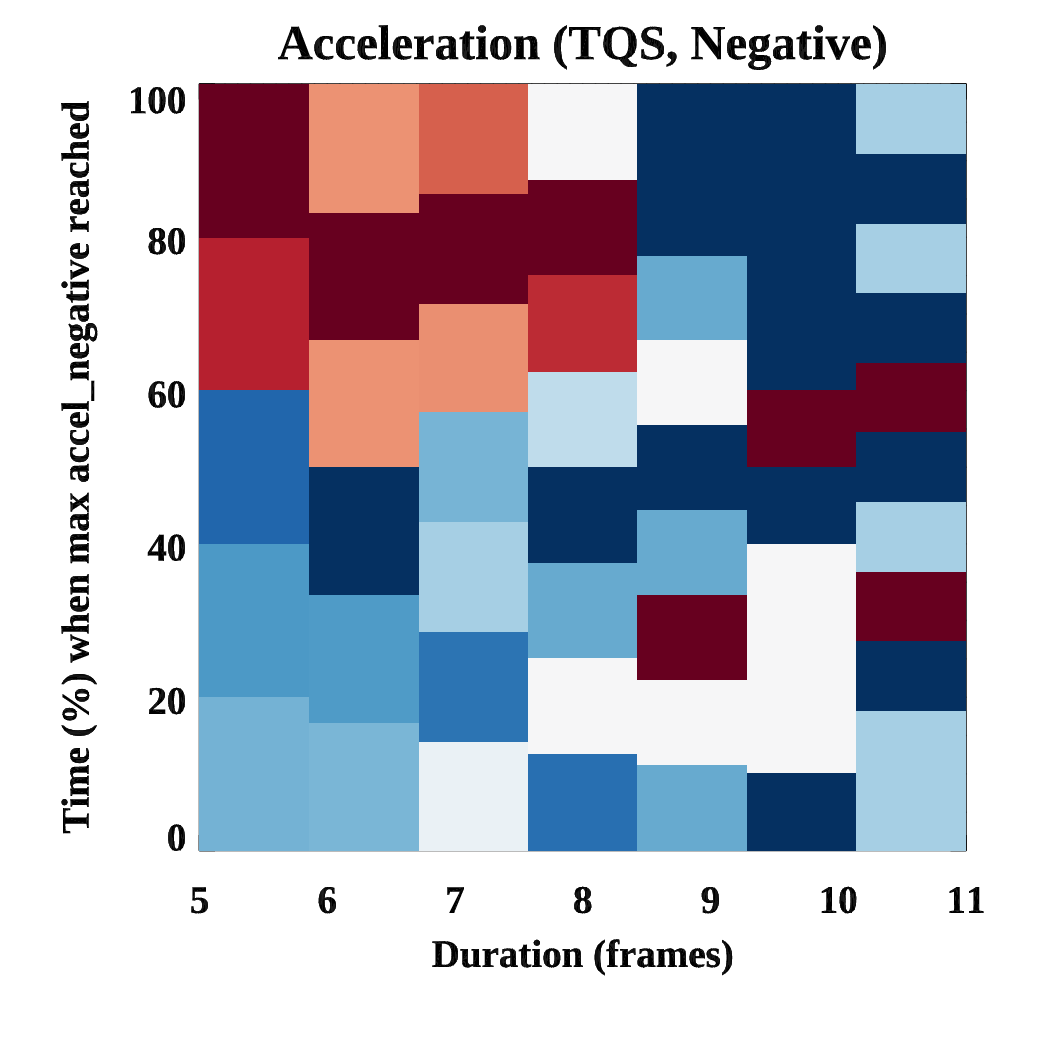}
    \caption{2D histogram plots of the temporal position (\%) at which (left column) positive acceleration and (right column) negative acceleration reach their maximum values
vs BP duration (in frames). Top middle and bottom rows represent results for all BPs, AQS BPs and TQS BPs, respectively.}
    \label{fig:pos_neg_2D}
\end{figure}

Key differences between AQS and TQS BP evolution are noted:

\begin{itemize}
    \item SL AQS BPs can reach their maximum total brightness at any point during their life (expect the beginning or the end), with a slight preference for doing so near the beginning of their life
    \item Long-lived (LL) AQS BPs are more likely to reach maximum total brightness near the beginning of their life, although the results are somewhat scattered
    \item SL BPs, regardless of whether they are AQS or TQS, can reach their maximum intrinsic brightness at any point in their lifetime (again excluding the start and end)
    \item LL BPs (both AQS and TQS) are more likely to reach maximum intrinsic brightness near the beginning of their lifetime (with some scatter in AQS results)
    \item SL BPs (both AQS and TQS) most commonly reach maximum POS speed around the midpoint of their life. However, this pattern becomes more dispersed with increasing BP lifetime
    \item SL BPs (both AQS and TQS) can reach maximum POS acceleration at any stage of their life except at the midpoint. The small peak at the 50th percentile in Figure \ref{fig:max_temp_pos}d likely corresponds to the small contributions from the darkest bins in Figure \ref{fig:attribute_vs_dur}i and \ref{fig:attribute_vs_dur}n near the 50\% mark
    \item All BPs generally show a tendency to reach maximum acceleration either early or late in their lifetime. This trend appears more scattered in AQS BPs, likely due to the smaller number of AQS observations
    \item LL TQS BPs show a clear tendency to reach their maximum acceleration later in life, typically around the 80th percentile
    \item BPs tend to reach their maximum area near either the beginning or the end of their lifetime (20th and 80th percentiles), and are unlikely to do so at the midpoint or precisely at the very start or finish
    \item However, SL BPs (both AQS and TQS) can achieve their area maxima at any point in their lives (except the beginning and end)
\end{itemize}

\subsection{Interpretation}

The fact that POS speeds rarely peak near the start or end of a BP’s lifetime suggests that BPs follow an arched vertical trajectory along current sheets, moving from one magnetic footpoint to another. Figure \ref{fig:schem_1} provides schematic representations of these BP motion along current sheets, viewed from the side. The yellow box at the bottom denotes the photosphere. This would naturally result in the highest POS speed and travel distance occurring at the apex of the trajectory - as is suggested by the relative sizes of both arrows and green lines in figure \ref{fig:schem_1} - approximately halfway through the BP’s lifetime, while the footpoints would be the least likely locations for maximum POS speed.

This interpretation is consistent with previous findings that AQS BPs tend to travel at lower POS speeds, as well as with the idea that POS speeds are influenced by magnetic field orientations. While magnetic fields in active domains are likely more vertically inclined, this would simply result in a steeper arch in the BP’s motion rather than contradicting the conclusions of \cite{Humphries_2024}, as can be seen in the left-hand portion of figure \ref{fig:schem_1}. This also aligns with the lack of significant differences between the POS acceleration profiles of active and quiet BPs, suggesting that POS acceleration is not necessarily correlated with LOS acceleration.

If BPs were solely the result of energetic eruptions, such as chrompsheric magnetic reconnection or flare-like events, one would expect their brightness and size to peak near the beginning of their lifetime. However, figures \ref{fig:max_temp_pos}(a), (b), and (e) show otherwise, demonstrating that maximum brightness and area can occur at almost any point. Additionally, figure \ref{fig:max_temp_pos}(d) shows large peaks around the 30th, and 70th percentiles with potentially a smaller peak near the 50th percentile, suggesting that BPs undergo significant POS acceleration just before, during (although less significantly), and after the midpoint of their lifetime. This further supports the idea that BPs follow curved magnetic loops, experiencing acceleration as they traverse different segments of the loop. 

Accounting for the ideal geometry of the loops suggested in figure \ref{fig:schem_1} as well as the vector diagram therein (seen in purple), if the POS acceleration is dependent on the angle from the photosphere at which the BP moves along the loop ($\theta$), then the following equations describe such acceleration:

\[V_{POS}=V\cos{\theta}\]
\[a_{POS}=\frac{d}{d\theta} V_{POS}\]
\begin{equation}
    \therefore a_{POS}=-V\sin{\theta}
    \label{eq_1}
\end{equation}

As a result, BPs should experience their smallest POS acceleration at the loop apex, while their greatest POS accelerations occur at the beginning and end of their lifetime. 
The large range of likelhood to reach maximum acceleration at both ends of BP lifetimes may correspond to the suggestion that, as figure \ref{fig:schem_1} demonstrates, the footpoints are located within the photosphere, and that the histograms are the result of BP detections only after they reach the chromopshere; if it were possible to detect the BP as it is emitted from or returns to a photospheric footpoint, then the acceleration peaks would likely be closer to the 0th and 100th percentiles. 

While the schematic paths appear relatively simple in a side-on perspective, BP paths are likely far more intricate. Previous findings by \cite{Humphries_2024} demonstrate that both active and quiet BPs can exhibit complex POS trajectories, likely reflecting the turbulent and chaotic nature of the chromosphere. Such complexity may also contribute to the variability observed in figure \ref{fig:max_temp_pos}(c), where BPs can reach their maximum speed at points other than the midpoint of their loop traversal. These variations in speed as well as those of acceleration may be influenced by differences in plasma density and sound speed at varying heights. Depending on the altitude of a given magnetic loop, the local plasma properties at the apex may differ significantly from those elsewhere within the loop. Notably, BP speeds near the apex appear to exceed the typical $\sim8$ kms$^{-1}$ sound velocity of the low chromosphere \citep{Anderson_1989}. \cite{Molnar_2021}] demonstrate that acoustic waves propagating into higher layers of the solar atmosphere encounter rapidly decreasing plasma densities, while the temperature - and thus the sound speed - remains approximately constant. This implies that BPs can reach POS speeds near the arch apex that exceed chromospheric sound speeds by up to a factor of $\sim8$.

Additionally, the range of possible percentiles at which total brightness and area may reach their maximum may also be due to variations in chromospheric density. Total brightness peaks are unlikely to be the result of LOS intensity bias (e.g., due to an elongation of BPs) as this would result in an inverse relationship between total brightness peaks and area peaks, which is not the case according to figures \ref{fig:max_temp_pos}a and \ref{fig:max_temp_pos}e. 

\begin{figure*}[t]
    \centering
    (a)\includegraphics[width=0.45\textwidth]{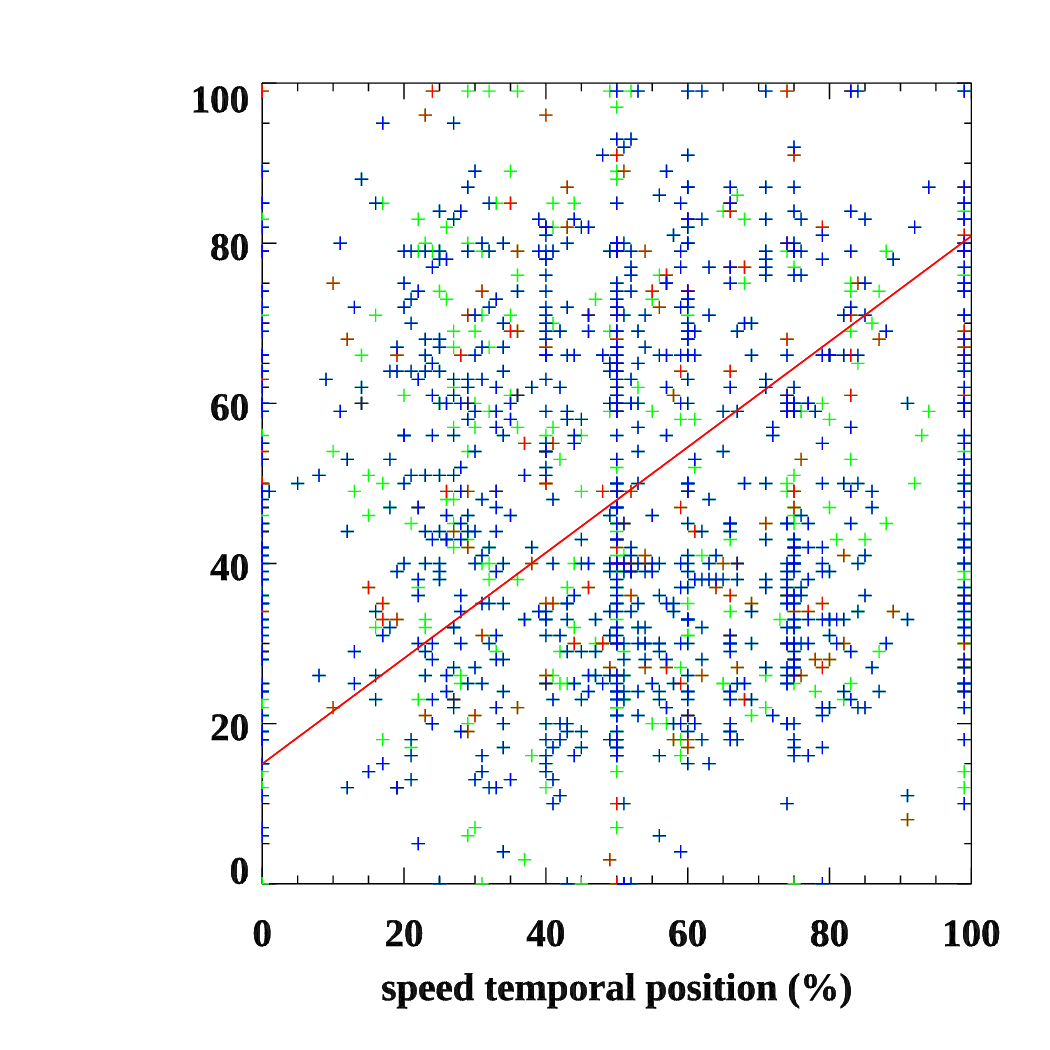}
    (b)\includegraphics[width=0.45\textwidth]{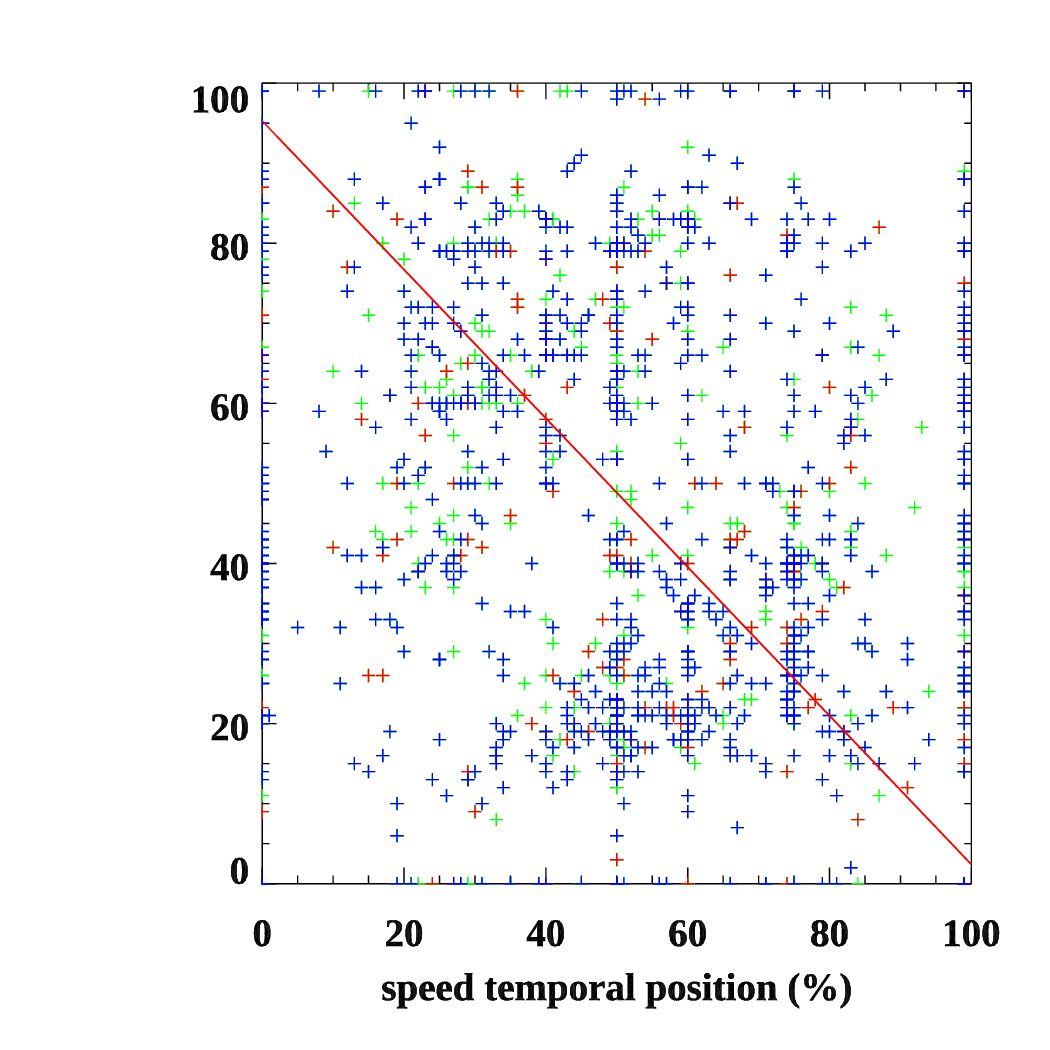}
    \caption{Scatter plots of the temporal position at which BPs reach their maximum speed vs (a) the temporal position at which BPs reach their maximum area and (b) the temporal position at which BPs reach their maximum intrinsic brightness. The red lines are the result of a robust deming fit.}
    \label{fig:scatter}
\end{figure*}

Nevertheless, \cite{Humphries_2024} also demonstrates that a great majority of BPs follow mostly straight POS paths of motion (based on the ratio of 'overall' travel distance to total travel distance). The idealised loop geometry, therefore, is an acceptable approximation for explaining the evolution of the majority of BPs, and that BPs that deviate significantly from straight-line paths correspond to the results that do not match with idealised loop geometry. 

We propose that, considering the thermal range of IRIS's 1400 \AA\ channel, BPs do not originate within the chromosphere itself but are instead the result of magnetic reconnection occurring in the underlying photosphere. The observed BPs could represent either a flow of plasma or an energy transfer along current sheets, propagating into the chromosphere along closed magnetic loops, typically following non-linear POS trajectories as suggested previously. In particular, based on BPs speeds frequently exceeding that of the chromosphere, energy transport via wavefronts — such as shockwaves — remains plausible.

Unlike speed, size and brightness can peak at nearly any point in a BP’s lifetime, though they are least likely to do so within the first and final 10\% of their lifespan. One possibility is that, under certain chromospheric conditions, BPs experience localized energy releases or trigger surrounding plasma, leading to increases in brightness and detected area. However, given the turbulent and chaotic nature of the chromosphere, these conditions may arise unpredictably throughout a BP’s evolution. Interestingly, both AQS and TQS BPs appear less likely to reach both their total and intrinsic brightness as well as area at the moment they are traveling fastest. This can be seen in figure \ref{fig:max_temp_pos}a, \ref{fig:max_temp_pos}b and \ref{fig:max_temp_pos}e when compared to figure \ref{fig:max_temp_pos}d.

\subsection{Acceleration peaks}

Separating the acceleration results of figures \ref{fig:max_temp_pos}d, \ref{fig:attribute_vs_dur}d, \ref{fig:attribute_vs_dur}i, and \ref{fig:attribute_vs_dur}n into their positive and negative constituents yields those presented in figures \ref{fig:pos_neg_1D} and \ref{fig:pos_neg_2D}. Figure \ref{fig:pos_neg_1D} demonstrates the likelihood of (a) positive and (b) negative accelerations reaching their maximum values in the same fashion as figure \ref{fig:max_temp_pos}, whereby red and blue represent AQS and TQS BPs, respectively. Figure \ref{fig:pos_neg_2D} is constructed in the same fashion as figure \ref{fig:attribute_vs_dur}, whereby the left column represents Positive accelerations, the right column represents negative accelerations, (a)-(b) represents results from all BPs, (c)-(d) represent AQS BP results, and (e)-(f) represent TQS BP results. While the preference for the either side of the 50th percentile remain for both positive and negative acceleration, regardless of whether the BPs are AQS or TQS, it is clear that positive acceleration and negative acceleration differentially occur at the beginning and end of BP lifetimes, respectively. This supports with the suggested magnetic loop geometry whereby, according to equation \ref{eq_1}, POS acceleration is positive and negative at the beginning and end of a BP's lifetime, respectively.

\subsection{Scatter Plots}

Further insights can be drawn from the preliminary scatter plot results in figure \ref{fig:scatter}, which compare the temporal position at which BPs reach their maximum speed to (a) their maximum area and (b) their maximum intrinsic brightness. Red and blue crosses indicate AQS and TQS BPs, respectively, while green crosses indicate BPs that are neither AQS or TQS. No clear correlation is seen in (a), indicating that size evolution is independent of speed evolution. Conversely, (b) reveals two distinct clusters in the top-left and bottom-right of the plot, separated by a gap. This suggests that BPs that reach their peak speed early in their lifetime tend to achieve their highest intrinsic brightness later, while those that reach maximum speed later in their lifetime tend to brighten earlier. It is also apparent that the AQS/TQS distinction is inconsequential to the results of these two BP clusters. The underlying cause of this behaviour remains unclear, but combining this with the peaks from figure \ref{fig:max_temp_pos} buttresses the potential for two distinct classes of BPs.

\subsection{Photospheric Reconnection in Similar Phenomena}

The characteristics of magnetic reconnection depend on the atmospheric layer in which it occurs. The Sweet-Parker model suggests that reconnection in the photosphere is largely isothermal, while reconnection in the transition region (TR) and upper chromosphere leads to significant plasma heating \cite{Litvinenko_2007}. Perhaps observations with additional wavelength can help determine whether BPs exhibit plasma heating, as prior studies suggest that these BPs are indeed multi-thermal in nature \citep{Humphries_2021_b}. 

\cite{Takasao_2013} find that slow-mode shock waves generated by magnetic reconnection in both the photosphere and chromosphere can result in the acceleration of chromospheric jets. They determined that reconnection near the photosphere leads to plasma moving along magnetic field lines, generating slow shock waves that lift the TR, whereas reconnection in the upper chromosphere results in the acceleration of plasma via Lorentz forces and abrupt magnetic fields motions. While chromopsheric jets are unlike BPs, the former scenario typically results in the emergence of cold plasma jets, while the latter results in emission of hot plasma jets. These BPs exhibit as brightenings in IRIS's 1400 \AA\ Si {\sc{iv}} channel, which strengthens the suggestion that, if reconnection is indeed involved, that it occurs within the photosphere.

Observations of Ellerman bombs (EBs) further support the notion that photospheric reconnection may drive small-scale brightenings. \cite{Pariat_2004} and \cite{Watanabe_2011} find that EBs result from undulatory flux emergence, where photospheric reconnection interacts with existing vertical network fields. While BPs do not exhibit the repetitive flaring seen in EBs, examining BP brightness evolution over their lifetime may reveal a pattern indicative of resistive flux emergence. However, such an analysis would necessitate IRIS data with very short cadences. Furthermore, \cite{Young_2018} suggests that UV bursts arise from small-scale reconnection in the upper photosphere or lower chromosphere. They demonstrate that UV bursts exhibit small proper motions and track photospheric magnetic dynamics, as opposed to traveling along extended loops structures. However, being of a similar size and duration to that of UV bursts, the evolution of BPs could still be linked to reconnection in the lower atmosphere. \cite{Young_2018} also notes that UV bursts display dynamics associated with currents sheets, such as fast-moving, dense plasmoids, or turbulence induced in the surrounding plasma by outflow jets associated with reconnection events. This is similar to \cite{Humphries_2024}'s suggestion that BP motions are linked to current sheets, and that the earlier suggestion that turbulence in the surrounding plasma may have some effect on their evolution. Similarly, \cite{Kahil_2022} concluded that most campfires - which occur 1000-5000km above the photosphere - originate from magnetic reconnection at their photospheric footpoints, although some are driven by alternative heating mechanisms, such as magnetic flux cancellation \citep{Chae_1998,Panesar_2021}.

\section{Conclusions} \label{sec:con}

We present the results of our BP lifetime analysis as well as a comparative analysis between ''Active" and ''Quiet" BPs within a QS region. Our study assesses $\sim1300$ BPs, providing insight into their dynamical evolution and physical characteristics. This work specifically examines when BPs reach the maximum values of their various attributes. We extract and plot data for maximum brightness, intrinsic brightness, POS speeds (and travel distance), POS acceleration, and area.

Our findings indicate that BPs can reach their peak brightness (both total and intrinsic) and maximum size at nearly any point in their lifetime, except during the first and final 10th percentile. We determine that Active and Quiet BPs exhibit similar trends in each attribute. Both Active and Quiet BPs are most likely to reach their maximum speed at the halfway point of their lifetime with steady drop-offs of likelihood on either side of this peak. POS acceleration occurring predominantly near the beginning and end of BP lifetimes.

We interpret these findings as evidence that BPs, whether viewed as plasma movements or energy transfers, travel along current sheets that follow an arched trajectory. The midpoint of their lifetime likely corresponds to the crest of this arch, which facilitates the greatest POS speeds. If BPs originate from magnetic reconnection, it is likely that this reconnection occurs within the photosphere at some footpoint. Following the emergence of the BP into the chromosphere (whereby it becomes visible in the 1400 \AA\ channel), they travel along loop structures before returning to another footpoint within the photosphere. The fact that BP size and area can peak at almost any point in their lifetime suggests that complex interactions with the surrounding chromosphere play a significant role in their evolution.

The presence of two distinct regimes of BPs is speculatively suggested, whereby one group reaches its maximum speed early in its lifetime but is unlikely to reach its maximum intrinsic brightness during the same period, and another group that attains peak speed later in its lifetime while also being unlikely to reach maximum intrinsic brightness during that stage. More research is needed to understand how the emerging plasma or energy transfer interacts with the chromosphere as it follows loop structures.

Speculatively, \cite{Fischer_2017}'s analysis of exploding granule structures observes similar physical effects to those of our proposed BP motions and evolution, whereby rapidly expanding granules result in horizontal flows of material, while trapped material in bordering intergranular lanes result in upward-traveling shockwaves. These differential flows of material correlate with our interpretation of BP motion as well as those suggested in \cite{Humphries_2024}. BPs are also of a similar size to granules - 400km radius vs 1000km radius, respectively - but are much shorter lived -$\sim180$s vs $\sim600-900$s \citep{Ellerwarth_2021}, respectively. However, it is reasonable that BPs would be shorter lived than the granules from which they potentially emerge, whereby granules may emmit several BPs during their own lifetime along multiple POS directions. 

As a next step, we aim to expand this study by including an active region dataset for direct comparison. Additionally, analyzing another QS dataset will help determine whether our findings remain consistent across different latitudes, longitudes, and solar cycles. Future work will also involve evaluating the behavior of BP in the 1330 \AA\ and 2796 \AA\ channels. Given the discrepancies in temporal and spatial results from previous multi-wavelength studies \citep{Humphries_2021_b} - particularly concerning the cooler 2796 \AA\ channel - a detailed comparison of BP evolution at different temperatures and atmospheric heights will be invaluable for understanding their role in solar atmospheric dynamics.\\

\begin{acknowledgments}
    L.H. and D.K. acknowledge the Science and Technology Facilities Council (STFC) grant ST/W000865/1 to Aberystwyth University. H.M. acknowledges the STFC grant ST/S000518/1 to Aberystwyth University. D.K. acknowledges the Georgian Shota Rustaveli National Science Foundation project FR-22-7506. L. H. acknowledges the Worshipful Livery Company of Wales award, Gwobr Bryan Marsh Gold Development, Runner-Up. The SDO and HMI data used in this work are courtesy of NASA/SDO and the AIA, EVE, and HMI science teams. IRIS is a NASA small-explorer mission developed and operated by LMSAL with mission operations executed at the NASA Ames Research center and major contributions to downlink communications funded by ESA and the Norwegian Space Centre. The authors wish to acknowledge FBAPS, Aberystwyth University, for providing computing facilities and support.
\end{acknowledgments}

\vspace{5mm}

\facilities{Aberystwyth University, IRIS
            }

\software{IDL
          }
\bibliography{paper5}
\bibliographystyle{aasjournal}

\section{Appendix}

BPs have a range of lifetime durations, quantized to an integer number of observed time steps. Each brightening has a recorded time step when some characteristic (e.g., brightness, or speed) reaches a maximum value. This appendix describes how we visualise this information in a histogram while avoiding biases due to the distribution of different number of time steps. 

\begin{figure}[]
\begin{center}
\includegraphics[width=12.0cm]{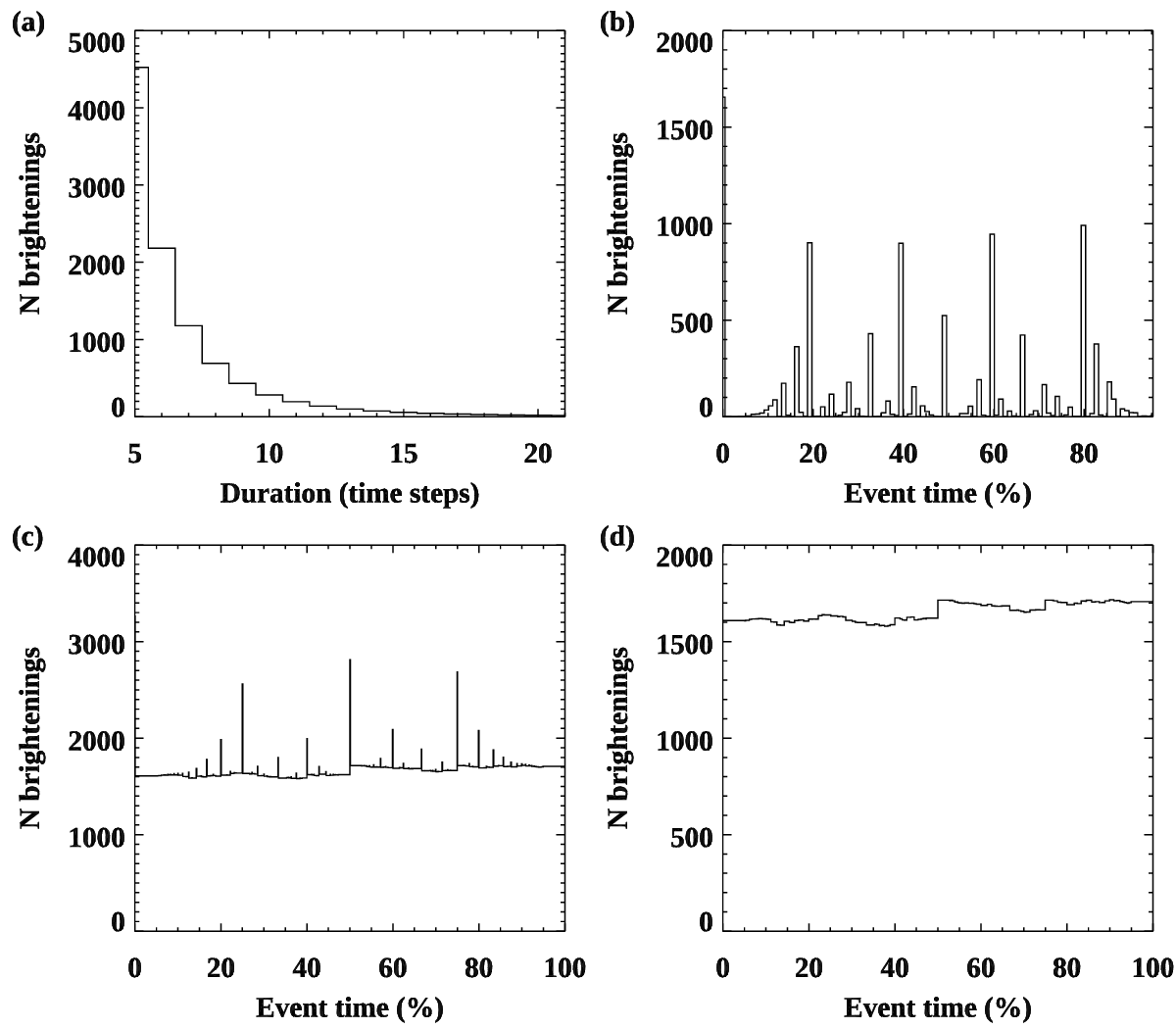}
\end{center}
\caption{(a) The distribution of duration for a set of 10,000 synthetic BPs. This distribution is approximately similar to what we find in real data. (b) Histogram of when an event occurs during a BP's lifetime, as percentage of the total duration of each BP. (c) Raw irregular-binsize histogram where the duration, or number of time steps, of each BP is taken into account. (d) Filtered irregular-binsize histogram (3-bin width median smoothing applied to the histogram of (c)).}
\label{fighist}
\end{figure}

To demonstrate and validate the approach, we use synthetic data for 10,000 brightenings, and the distribution of these BPs's durations, in time steps, is shown in Figure \ref{fighist}a. This distribution is similar to real data, where shorter durations contain larger populations. For these BPs, we assign to each an 'event' at one time step during its lifetime, selected at random. For real data, this would be where a maximum value (e.g., brightness, or speed) is reached. The random time-step index is converted to percentage of a BP's lifetime. For example, for a BP with 5 time steps, we record a random event at time step 2, and this converts to a percentage of duration of 40\%. Figure \ref{fighist}b shows a standard histogram of all event times using 100 bins. This distribution is biased by the distribution of brightening durations. For example, there are five distinct peaks at 0,20,40,60 and 80\%, formed by the BPs with 5-time-step durations. The peak at 0\%\ is highest, since all durations can contribute to this bin.  

To avoid this bias, we must form separate histograms for each duration, with the number of bins for each histogram equal to the number of time steps. We achieve this by setting a large number of bins (1000). For a duration length of, e.g., 5 time steps, we split the number of histogram bins into 5 equal portions. Then the histogram value spanning each portion is given by the number of events falling within that bin. So for an event at time step 2, we increase the histogram between bins 400 and 600 by 1. This is repeated for all events, then repeated for all durations. This gives the histogram shown in figure \ref{fighist}c. The narrow spikes arise due to the quantisation of durations, and narrow overlaps between the distribution of bins. We remove these spikes by median smoothing over 3 bins, with the final result shown in figure \ref{fighist}d. This is an uniform histogram as we expect from the random distribution of events.

\end{document}